\documentclass[seceq,supplement]{ptptex}
\usepackage{amsmath,amssymb,bm}
\usepackage[dvips]{graphicx}
\usepackage{subfigure,wrapfig} 
\usepackage{color}

\newcommand{\6}{\partial}

\newcommand{\Lie}{\pounds}

\newcommand{\m}{\mu}
\newcommand{\n}{\nu}
\newcommand{\la}{\lambda}
\newcommand{\La}{\Lambda}

\newcommand{\ro}{\rho}
\newcommand{\s}{\sigma}
\newcommand{\de}{\delta}

\newcommand{\ga}{\gamma}

\newcommand{\no}{\notag \\}

\newcommand{\poincare}{Poincar\'{e} }

\newcommand{\na}{\nabla}

\newcommand{\DDR}{\overset{\text{\tiny $(\ga)$}}{R}}

\newcommand{\pj}[2]{\ga_{#1}^{~#2}}

\newcommand{\T}{\mathcal{T}}
\newcommand{\bfga}{\boldsymbol{\ga}}
\newcommand{\bfg}{\boldsymbol{g}}
\newcommand{\Msolar}{M_\odot}

\notypesetlogo
\title{
Black Holes in Braneworld Models}

\author{Norihiro \textsc{Tanahashi}$^{1}$ and Takahiro \textsc{Tanaka}$^{2}$}

\inst{$^1$Department of Physics, University of California, Davis, USA\\
$^2$Yukawa Institute for Theoretical Physics, Kyoto University, \\
Kyoto 606-8502, Japan}

\abst{In this review, we summarize current understandings of black
hole solutions in various braneworld models, including the 
Arkani-Hamed-Dimopoulos-Dvali model, the Randall-Sundrum (RS) models, 
Karch-Randall (KR) model and the Dvali-Gabadadze-Porrati model. 
After illustrating basic properties of each braneworld model, we introduce the
bulk/brane correspondence in the RS and KR braneworld models, adding 
supporting evidence for it. We then summarize the studies on
braneworld black hole solutions, which consist of constructing exact 
or approximate solutions and investigating the phase diagram of
solutions. In the study of phase diagram, we will also expound the 
implications of the bulk/brane correspondence to the braneworld black holes.
}

\begin{document}
\maketitle

\tableofcontents

\section{Introduction}
\label{Sec:intro}

Contemporary candidates of quantum gravity, such as string theory or M-theory,
predict that our spacetime is higher-dimensional even though we observe seemingly
four-dimensional world.
To reconcile this apparent conflict in the spacetime dimensionality, we have to 
introduce some mechanisms into the spacetime structure.
One of such mechanisms is the Kaluza-Klein (KK)
compactification~\cite{Kaluza:1921tu,Klein:1926tv},
in which extra dimensions are manifestly compactified into a finite size manifold. 
Another one is the braneworld model, which we will focus on in this review.

A braneworld models is composed of higher-dimensional bulk spacetime and
lower-dimensional brane in it. 
Standard model particles are confined onto
the brane but gravity is allowed to propagate in the bulk.
These models are partly motivated by string theory in
the following sense.
In string theory, membrane-like solutions emerge as a result of string condensation. 
Open strings attached on this membrane will behave as matter
fields confined on it. Closed strings, on the other hand, can freely move away from the
brane, and will behave as graviton propagating in the bulk.
The braneworld models will capture gravitational aspect of such membrane-like
solutions.

Since graviton propagates in the bulk and then the gravity in this model is
higher-dimensional, we need some mechanisms 
to recover the four-dimensional gravity on the brane.
The simplest way would be cutting off the extra dimensions in a 
manner similar to the Kaluza-Klein compactification.
In the Arkani-Hamed-Dimopoulos-Dvali (ADD)
model~\cite{ArkaniHamed:1998rs,ArkaniHamed:1998nn},
the first braneworld model, 
and in the Randall-Sundrum I (RS-I) model~\cite{RS-I},
the four-dimensional gravity is recovered in this way.
Second way is to introduce the so-called warped compactification, 
with which we can localize the gravity onto the brane even when
the bulk extends infinitely.
This mechanism is used in
the Randall-Sundrum II (RS-II) model~\cite{Randall:1999vf}, 
which is composed of
negatively curved bulk spacetime and a four-dimensional brane with tension.
Yet another way to localize gravity is to manifestly
introduce the four-dimensional Einstein-Hilbert action localized 
on the brane, which is adopted in the Dvali-Gabadadze-Porrati (DGP) 
model~\cite{DGP}.

These gravity localization mechanisms can be assessed by perturbation analyses
in the weak gravity regime, and indeed they are shown to work well. 
In the strong gravity regime, however,
it is not clear how the gravity behaves due to complexity of
the Einstein equations.
Especially, the properties of black hole solutions in braneworld models are not
yet fully understood though much effort has been devoted to this issue.
In this chapter, we will review current understanding about such 
black hole solutions in braneworld models.

As for the RS braneworld model, there is another noteworthy point.
Since its unperturbed bulk spacetime is anti-de Sitter (AdS) space, 
it is expected that the AdS/CFT correspondence~\cite{Mal,Aharony:1999ti} 
applies also to this model~\cite{WittenComment,VerlindeComment}. 
The AdS/CFT correspondence indicates the duality between classical gravity in the 
bulk and conformal field theory on the AdS boundary. 
A novel feature of the duality in the RS model is that the quantum theory 
on the boundary couples to gravity, while they are decoupled in the original
AdS/CFT correspondence.
Even though there is no direct proof for this modified duality so far,
a vast number of works are supporting 
it.
This bulk/brane correspondence in braneworld models will be a topic 
that we focus on in this review.

This chapter is organized as follows.
We firstly review in \S\ref{Sec:overview} the basics of braneworld models, 
such as their geometrical structure 
and properties of weak gravity on it.
In \S\ref{Sec:Hol}, we 
introduce 
the effective Einstein equations on the brane and
the bulk/brane correspondence in the RS braneworld model. 
Next, we discuss black holes in the ADD model in \S\ref{Sec:ADDBH}, 
and those in the RS model in \S\ref{Sec:arguments}, 
mainly focusing on brane-localized black holes. 
Since
even a numerical solution describing a large brane-localized black hole 
in the RS model has not been found so far,
we begin with the predictions derived from the bulk/brane
correspondence.
After that,
we summarize the progress in studies on the brane-localized black 
holes and give arguments supporting the predictions.
We will also discuss
the studies on black hole solutions in other braneworld models, including 
the Karch-Randall (KR) model and the DGP model, in \S\ref{Sec:other}. 
Finally, we will briefly summarize this review and mention some 
prospects in \S\ref{Sec:summary}. 

\subsection*{Notations and Conventions}
In this chapter,
we express
the metric in the bulk as $g_{AB}$ with indices $A, B,\dots$, and
covariant derivative with respect to $g_{AB}$ as $\na_A$.
Correspondingly, we express
the metric on the brane and covariant derivative with respect 
to it as $\ga_{\m\n}$ and $D_\m$ 
with indices $\m,\n,\dots$\,.
We use indices $i,j,\dots$ for coordinates in the spatial direction when
necessary.

\section{Braneworld models}
\label{Sec:overview}

Braneworld models are higher-dimensional spacetime models 
in which the standard model particles are 
confined on a four-dimensional hypersurface (brane) while gravity can freely 
propagate in the higher-dimensional spacetime (bulk) which encompass the brane.
The first braneworld model is the ADD model
with tensionless brane and
compactified flat extra dimensions.
Later, this model is extended to the RS model that consists of 
non-zero tension branes and a warped,
i.e., negatively curved bulk spacetime.
Another variation of braneworld model is to include
the Einstein-Hilbert action localized on the brane, which is 
called the DGP model.
This model leads to gravity modification in a long distance scale while keeping 
the ordinary Einstein gravity in a short distance scale.

In this section, we introduce these braneworld models and explain 
the behaviors of weak gravity in these models.
Starting with a brief introduction of the ADD model in \S\ref{Sec:ADDbasic}, we
explain the RS model in \S\ref{Sec:RSbasic}.
We also introduce a variant of the RS model with an asymptotically AdS brane, 
which is called the KR model, in \S\ref{Sec:KRbasic}. 
Finally, we mention the DGP model in \S\ref{Sec:DGPbasic}.

\subsection{ADD model}
\label{Sec:ADDbasic}

The prototype of current braneworld models is the ADD 
model~\cite{ArkaniHamed:1998rs,ArkaniHamed:1998nn}, which is composed of 
a tensionless brane and flat compactified extra dimensions. This model was 
introduced to resolve the hierarchy problem between the Planck scale 
$M_\text{Pl}\sim 10^{19}\,\mathrm{GeV}$ and the electroweak scale 
$M_\text{EW}\sim 1\,\mathrm{TeV}$
(see Refs.~\citen{Kribs:2006mq} and \citen{Cheng:2010pt} 
about particle physics studies in this model,
and Ref.~\citen{Antoniadis:1998ig} 
about embedding of the model into string theory).

The action of gravity in $(4+d)$-dimensions is 
assumed to be simply given by Einstein-Hilbert action
\begin{equation}
 S={1\over 16\pi G_{4+d}}
       \int d^{4}x\int d^{d}y \sqrt{-g}\, R~, 
\end{equation}
where $R$ is the curvature scalar in 
$(4+d)$-dimensions. 
As in the case of Kaluza-Klein compactification, 
modes which have non-trivial structure in the directions 
of extra dimensions are massive. Hence, they are 
difficult to excite. Neglecting these massive excitations, 
we assume that the configuration is homogeneous 
in $y$-directions. Under this assumption, we can 
integrate over extra dimensions to obtain 
\begin{equation}
 S\approx {L^d \over 16\pi G_{4+d}}
      \int d^4x \sqrt{-g^{(4)}}\, R^{(4)}~, 
\label{Eq:reduce4ddaction}
\end{equation}
where $g^{(4)}_{\mu\nu}$ is the metric on a 
four-dimensional section, and $R^{(4)}$ is scalar 
curvature computed from $g^{(4)}_{\mu\nu}$. 
$L^d$ is the $d$-dimensional volume of extra dimensions. 

Comparing the reduced action (\ref{Eq:reduce4ddaction}) 
with the four-dimensional Einstein-Hilbert 
action,  
$S_{4D} = {1\over 16\pi G_4}$
$\int d^4x \sqrt{-g^{(4)}} R^{(4)}$, 
we find that the effective four-dimensional Planck 
mass $M_4 = G_4^{-1/2}$ is related to the 
original $(4+d)$-dimensional one 
$M_{4+d} = G_{4+d}^{-1/(2+d)}$
by 
\begin{equation}
 M_{pl}^2\approx M_{4+d}^{2+d} L^{d}~.
\label{ADDd}
\end{equation}
Then, for $M_4\sim 10^{19}\,\mathrm{GeV}$, $M_{4+d}$ is given as 
$
M_{4+d} \sim 10^{\frac{38-15d}{2+d}} \times 
\left(
\frac{0.1\mathrm{mm}}{L}
\right)^{\frac{d}{2+d}}
\mathrm{GeV}
$,
and $M_{4+d}$ becomes as low as TeV if we set $d=2$
and $L\sim 0.1\,\mathrm{mm}$.
Note that this relatively large extra dimension scale will not be probed by
high energy particle experiments since the standard model particles are confined
on the brane.
The model with $d=1$ is already excluded since $L$ must be astrophysical 
distance scale, while the models with 
$d\geq 3$ are observationally allowed for a wide range of model parameters. 
Since the brane is tensionless in this model, the gravitational property of this
model is similar to that of a KK compactified spacetime. Namely, long
distance behavior of gravity is similar to the four-dimensional one with the 
effective gravitational coupling mentioned above, while at a short
distance comparable to the size of extra dimensions the gravity becomes
higher-dimensional. 

\subsection{RS model}
\label{Sec:RSbasic}

The RS model is a braneworld model composed of branes with 
tension and a warped bulk spacetime with a negative cosmological constant.
The first model with a bulk bounded by 
positive and negative tension branes 
was introduced in Ref.~\citen{RS-I} 
to give an alternative viewpoint to the hierarchy problem.
Later, another variation of this model was proposed, in which 
the negative tension brane was sent to infinity~\cite{Randall:1999vf}. 
These models are called the RS-I and RS-II models, respectively.
We will explain basic features of gravity in these models below.

\subsubsection{RS-I model}

The unperturbed background of the RS-I model is composed of
a five-dimensional AdS bulk bounded by 
a positive tension brane located on the AdS boundary side 
and a negative tension brane on the other side. 
Its action is given by
\begin{eqnarray}
 S_\text{RS-I} & = & \frac{1}{16\pi G_5}
\int \!
d^5x \sqrt{-g} \left(
R - 2\Lambda_5
\right)
\cr
&&- \int \!
dy d^4x 
\sqrt{-\ga}
\left\{
(\la_+ +L_{m+}) \de\left(y\right)
+ (\la_- +L_{m-})\de\left(y-y_-\right)
\right\}~,
\end{eqnarray}
where 
$y$ is the spatial coordinate in the bulk direction, and
$\ga_{MN}$ is four-dimensional induced metric on 
$y=\text{constant}$ surfaces. 
The AdS boundary is at $y=-\infty$,
and the brane with positive and negative tension are located at $y=0$ and 
$y=y_->0$, respectively.
We impose $Z_2$-symmetry about the branes to the spacetime. 
In other words, we 
require that the spacetime has reflection symmetry about the branes.
$L_{\pm}$ is the Lagrangian of the matter
fields localized on the respective branes.  
The bulk spacetime is characterized by its curvature length $\ell$, which
is related to 
the five-dimensional cosmological constant $\Lambda_5$ by $\Lambda_5=-6/\ell^2$.
The brane tensions $\la_\pm$ are set to
\begin{equation}
 \la_+ = -\la_- = \frac{3}{4\pi G_5 \ell}
\end{equation}
so as to be compatible with 
a four-dimensional Minkowski spacetime on each brane. 
Then, the five-dimensional Einstein equations are given by
\begin{eqnarray}
R_{MN}-\frac{1}{2}Rg_{MN} & = & -\La_5 g_{MN} 
- 8\pi G_5 \Bigl\{
(\la_+ \ga_{MN}+\tau_{+MN})
\de\left(y\right)
\cr &&\qquad\qquad\qquad\qquad
+ (\la_- \ga_{MN}+\tau_{-MN})\de\left(y-y_-\right)
\Bigr\}~, 
\end{eqnarray}
where $\tau_{\pm MN}$ are the energy momentum tensors of the matter fields
localized on the branes. 
The delta functions on the right-hand side originate from the 
terms localized on the branes. 
Integrating this equation for an infinitesimal range across the brane, 
junction conditions at the brane are obtained as 
(see also \S\ref{Sec:SMS} for an alternative derivation)
\begin{equation}
\left[
K^\pm_{\m\n}-K^\pm\ga_{\m\n}
\right]^+_- = 8\pi G_5 (\la_{\pm} \ga_{\m\n}+\tau_{\pm\m\n})~,
\label{RS-IJC}
\end{equation}
where $K^\pm_{\m\n}\equiv -\pj{\m}{M}\pj{\n}{N}\na_M s^{\pm}_N$ is the extrinsic 
curvature and $s^M_\pm$ are the unit normal vectors 
of the respective branes. 
The notation $[f(y)]^+_-$ represents  
$f(y\to y_b+0)-f(y\to y_b-0)$ with $y_b$ being the 
location of the brane.
A simplest solution to these equations is given by
\begin{equation}
 ds^2 = 
e^{2|y-y_-|/\ell}
\eta_{\m\n}dx^\m dx^\n
+ dy^2~,
\label{RSImetric}
\end{equation}
where $\eta_{\m\n}$ is four-dimensional Minkowski metric.
Since the standard model particles are supposed to be confined on 
the negative tension brane, 
we normalized the warp factor $e^{2|y-y_-|/\ell}$ 
so as to be unity there. 

Notice that the above solution~(\ref{RSImetric}) remains to be a
solution even if the four-dimensional Minkowski metric $\eta_{\m\n}$ 
is replaced with any metric that solves the four-dimensional 
vacuum Einstein equations. 
Hence, substituting this simple metric form (the one with $\eta_{\m\nu}$
replaced with a generic four-dimensional metric), 
the five-dimensional gravitational 
action is reduced to an effective four-dimensional one. 
From this reduction, we can easily read that the effective four-dimensional Planck mass 
$M_4 = G_4^{-1/2}$ is related to the five-dimensional one $M_5 = G_5^{-1/3}$ as
\begin{equation}
 M_4^2 = \ell M_5^3 \left(
e^{2 y_-/\ell} - 1
\right)~.
\label{Mplrelation}
\end{equation}
If we set the brane position at $y_-/\ell \sim 37$, 
the energy scale in five dimensions are unified as 
$\ell^{-1}\sim M_5 \sim 1\,\mathrm{TeV}$.
Namely, we can absorb the energy hierarchy between the Planck scale and 
the electroweak scale into the geometrical 
hierarchy of $\mathcal{O}(10)$ in the RS-I model.
The particle phenomenology in this model was widely discussed~\cite{Kribs:2006mq}.

\subsubsection{RS-II model}

Let us take the limit of sending the negative tension brane to infinity
($y_-\to\infty$) in the RS-I model, 
and suppose that we are living on the positive tension brane.
Then, we obtain a model with one positive 
tension brane whose unperturbed metric is given by 
\begin{equation}
ds^2=
g^{(0)}_{MN} dx^M dx^N=
e^{-2|y|/\ell}
\eta_{\m\n}dx^\m dx^\n
+ dy^2
= \frac{\ell^2}{\left(|z|+\ell\right)^2}\left( \eta_{\m\n}dx^\m dx^\n + dz^2\right)~,
\label{RS-IImetric}
\end{equation}
\begin{wrapfigure}{l}{6.6cm}
\centering
\includegraphics[width=1.8cm,clip]{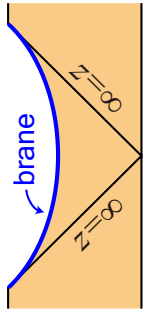}
\caption{%
Conformal diagram of the RS model.
}
\label{Fig:RS-IIconfD}
\end{wrapfigure}
where we introduced a conformal coordinate $z\equiv \text{sign}(y)\ell(e^{|y|/\ell}-1)$.
In this case the effective four-dimensional Planck mass is given  
by $ M_4^2 = \ell M_5^3$ (or equivalently $G_4=G_5/\ell$), and then 
this model is not useful to solve the hierarchy problem.


An interesting point of this model is that the higher-dimensional
gravity is effectively confined around the brane, 
although the extra dimension is infinitely extending.
The conformal diagram of this spacetime is depicted in Fig.~\ref{Fig:RS-IIconfD}. 
Thanks to the warping of the bulk spacetime and 
so-called volcano type gravitational potential 
due to the brane,
which is given by Eq.~(\ref{Eq:volcano}) below, 
the five-dimensional gravity is localized on 
the brane and an approximate four-dimensional gravity is realized on it.
This perturbative analysis on the gravity in the RS model
was done in Ref.~\citen{Garriga:1999yh}.

It is convenient to work in the so-called RS gauge, which is defined for the metric 
perturbation 
$h_{MN}\equiv g_{MN}-g^\text{(0)}_{MN}$ 
by 
\begin{equation}
h_{zz}=0= h_{\m z}\,, \quad
h_\m^{~\,\n}{}_{,\n}=0\,, \quad
h^\m_{~\m}=0\,.
\end{equation}
In this gauge, the equation of motion for the metric perturbation is 
simply given by
\begin{equation}
\left[
-\6_z^2 + V(z)\right] \psi_{\m\n}= 
\eta^{\ro\la}\6_\ro \6_\la
\psi_{\m\n}~,
\label{RSEoM}
\end{equation}
where  $\psi_{\m\n}\equiv \sqrt{|z|+\ell} \, h_{\m\n}$ and
\begin{equation}
V\left(z\right) 
= 
\frac{15}{4\left(\left|z\right|+\ell\right)^2} - 3 \ell^2 \de\left(z\right)~.
\label{Eq:volcano}
\end{equation}
Separating the variables as 
$\psi_{\m\n}\propto \hat u_m(z)e^{ik_\la x^\la}$, 
we obtain an eigenvalue equation
\begin{equation}
\left[
-\6_z^2 + V(z)
\right]\hat u_m(z)=m^2 \hat u_m(z)~,
\label{Eq:eigenvalueeq}
\end{equation}
where $m^2=-k_\m k^\m$ is the effective four-dimensional mass of the mode 
$\hat u_m(z)$.
The general solution that satisfies $Z_2$-symmetry about the brane is given in 
      terms of the Bessel functions as
\begin{equation}
\hat u_m(z)
=N_m\sqrt{|z|+\ell}\left[
J_1(m\ell)Y_2\bigl\{m(|z|+\ell)\bigr\}
-
Y_1(m\ell)J_2\bigl\{m(|z|+\ell)\bigr\}
\right]~,
\label{Bessel}
\end{equation}
where 
$N_m\equiv \left(m/2\right)^{1/2}\left\{J_1(m\ell)^2+Y_1(m\ell)^2\right\}^{-1/2}$ 
is the normalization constant
which is determined from the requirement 
$2\int_\ell^\infty \hat u_m(z) \hat u_{m'}(z)dz=\de(m-m')$.
Besides these modes, 
there is a discrete mass spectrum at $m^2=0$ with the wave function 
$\hat u_0(z)=\ell(|z|+\ell)^{-3/2}$, 
which is  called ``zero mode''. This zero mode wave function is nodeless. 
Since Eq.~(\ref{Eq:eigenvalueeq}) is an eigenvalue equation  
of Schr\"odinger type, 
the nodeless solution is the mode with the lowest eigenvalue. 
Therefore there is no bound state with $m^2<0$. 

From these mode functions, we obtain Green's function for 
$-\6_z^2+V(z)$
as
\begin{multline}
\hat G({\boldsymbol x},z,{\boldsymbol x'}, z') 
\\
\quad =
-
\!\!
\int_{-\infty}^{\infty}
\!\!
dt
\!\!
\int
\!\!
\frac{d^4 k}{(2\pi)^4}e^{ik_\m\left(x^\m-x'^\m\right)}
\!\!
\left[
\frac{
\hat u_0(z) \hat u_0(z')
}{{\boldsymbol k}^2-(\omega+i\epsilon)^2}
+ \int _0^\infty 
\!\!\!
dm\frac{
\hat u_m(z)\hat u_{m'}(z')
}{m^2+{\boldsymbol k}^2-(\omega+i\epsilon)^2}
\right],
\end{multline}
and the Green function for $h_{\m\n}$ is given by 
$G=\ell(|z|+\ell)^{-1/2}(|z'|+\ell)^{-1/2}\hat G$.
In the RS gauge, however, the brane moves from its original position when 
gravitational sources exist on the brane.
To obtain the induced metric on the brane, we have to 
transform the coordinates into the 
the ones in which the brane is located at $\bar y=0$, 
where bar ``$~\bar{}~$'' is associated for the distinction 
from the quantities in the original coordinates.
We further require $\bar h_{zz}=0=\bar h_{z\m}$, and 
we conventionally call them Gaussian-normal coordinates. 
Then, we obtain the induced metric components for a spherical source of mass $M$
on the brane as~\cite{Garriga:1999yh}
\begin{equation}
\bar h_{tt}=\frac{2G_4 M}{r}\left(
1+\frac{2\ell^2}{3r^2}
\right)~,
\qquad
\bar h_{ij}=\frac{2G_4 M}{r}\left(
1+\frac{\ell^2}{3r^2}
\right)\de_{ij}~.
\label{NewtonPotential}
\end{equation}
This $\bar h_{tt}$ is nothing but the modified 
Newton potential for the source $M$.
The metric perturbation in the bulk 
generated by a static source on the brane 
is summarized by~\cite{Garriga:1999yh} 
\begin{equation}
G({\boldsymbol x}, z, {\boldsymbol x'},0)\simeq
-\frac{\ell}{8\pi z^2}\frac{2r^2+3z^2}{\left(r^2+z^2\right)^{3/2}}~,
\label{GFbulk}
\end{equation}
which implies that the metric perturbation in the asymptotic region behaves as 
$\mathcal{O}(r^{-1})$ for a large $r$ and $\mathcal{O}(z^{-3})$ for a
large $z$.

%

\subsection{KR model}
\label{Sec:KRbasic}

We saw that the five-dimensional gravity is localized onto the brane in
the RS-II model due to the bulk spacetime warping.
Karch and Randall found that this feature persists 
even when the brane tension is detuned and the four-dimensional
spacetime on the brane becomes asymptotically AdS spacetime.
This model with asymptotically AdS branes was firstly proposed as an extension of 
the RS-II model in Ref.~\citen{KR}, and 
its embedding to the string theory was discussed 
later in Refs.~\citen{KRstring_orig,KRstring,DeWolfe,DeWolfe2}.
We briefly explain this KR braneworld model below.

Starting with the RS-II model,
we decrease the brane tension as
\begin{equation}
\la=\frac{3(1+\de)}{4\pi G_5 \ell }~,
\end{equation}
with $\de<0$.
Then, the unperturbed five-dimensional metric for this model becomes
\begin{gather}
ds^2 
=
 \frac{\ell^2}{L^2\sin^2\left\{
(|z|+z_0)/L
\right\}
}
\left(
\ga_{\m\n}dx^\m dx^\n + dz^2
\right)~,
\no
\ga_{\m\n}dx^\m dx^\n 
= 
-\left(1+\frac{r^2}{L^2}\right)
dt^2+ \frac{dr^2}{1+\frac{r^2}{L^2}} + r^2 d\Omega_\text{II}~,
\label{KRmetric}
\end{gather}
where $z_0\equiv L\arcsin(\ell/L)$ and the brane is located at $z=0$. 
The warp factor $\ell/L\sin\{(|z|+z_0)/L\}$ is chosen so that it 
becomes unity on the brane.
The coordinate $z$ varies in the range $0\leq |z|<\pi L-z_0$. 
An effective negative cosmological constant 
$\La_\text{4D}=-3/L^2 = 3(2\de+\de^2)/\ell^2$
arises on the brane, and hence the unperturbed induced metric on the brane 
is given by four-dimensional AdS space with the curvature length $L$.

For $|z|\ll z_0$, 
the warp factor in Eq.~(\ref{KRmetric}) is approximated as 
\begin{equation}
\frac{\ell}{L\sin\left\{
\left(|z|+z_0\right)/L
\right\}}
\simeq
\frac{\ell}{
l+\cos\left(z_0/L\right)|z|
}~,
\end{equation}
which is essentially the same as that of the RS-II 
model in Eq.~(\ref{RS-IImetric}).
Such a region with $|z|\ll z_0$ is responsible for 
four-dimensional gravity on the brane on scales shorter than $L$, 
thus we may expect that the confinement of the gravity also 
occurs in the KR model if we focus on such shorter scale.
This naive expectation can be confirmed by the analysis 
parallel to the RS-II case.
Let us consider the transverse-traceless perturbation of the background 
metric~(\ref{KRmetric}) defined by $h_{MN}\equiv g_{MN}-g^\text{(0)}_{MN}$
and $h_{zz}=0=h_{\m z}$, $D^\n h_{\n\m}= 0$ and $h^\m_{~\m}=0$,
where $D_\m$ is the covariant derivative with respect to $\ga_{\m\n}$.
Then, the linearized equations of motion become 
$\left[-\6_z^2 +V(z)\right]\psi_{\m\n}=m^2\psi_{\m\n}$ with
\begin{equation}
 V(z)=-\frac{9}{4L^2} + \frac{15}{4L^2\sin^2\left\{
\left(|z|+z_0\right)/L\right\}}
-\frac{3\sqrt{L^2-\ell^2}}{\ell L}\de(z)~,
\label{Eq:KRvolcano}
\end{equation}
where 
$\psi_{\m\n}\equiv \left[L\sin\{(|z|+z_0)/L\}\right]^{1/2}h_{\m\n}$.
The shape of the potential is quite different from the RS-II case. 
The potential decreases for $(|z|+z_0)/L<\pi/2$
as we move away from the brane, but it turns to increase for 
$(|z|+z_0)/L>\pi/2$.  
Hence, the spectrum of eigenvalues $m^2$ becomes totally discrete. 
If we neglect the presence of the boundary at $z=0$ 
(dropping the delta function in the left-hand side), we find that 
eigenvalues are given by $m^2=n(n+3)/L^2~(n=1,2,3,\cdots)$.
The delta function in Eq.~(\ref{Eq:KRvolcano}) imposes a boundary 
condition on $\psi_{\m\n}(z)$ at the brane, which is given by 
$\psi_{\m\n}'(z)=-3\sqrt{L^2-\ell^2}\psi_{\m\n}(z)/2\ell L$.
One finds that imposing this boundary condition does 
not change the above eigenvalues significantly. 
As in the RS-II model, $h_{\m\n}$ proportional to the squared warp factor 
satisfies the equations of motion with $m^2=0$, 
including the junction condition at $z=0$. 
However, in the present case, this function is divergent at 
$(|z|+z_0)/L=\pi$, and hence it is not normalizable. 
By imposing the normalizability, we find that the mass eigenvalue is 
shifted from zero. 

We present a way how to roughly estimate this mass shift. 
We consider the region $(|z|+z_0)/L \lesssim\pi/2$ and denote the $z$-dependent
part of the wave function 
as $\psi_+ +\delta\psi$, where $\psi_+$ is the function 
corresponding to $h_{\m\n}$ proportional to the squared warp factor,
i.e., $\psi_+\approx 
(|z|+z_0)^{-3/2}\hat\psi_+ $, and $\delta\psi$ is 
the deviation from it. 
The green function for $[-\6_z^2+V(z)]$ is given by
$G(z,z')\approx\left\{
\psi_+(z)\psi_-(z')-\psi_-(z)\psi_+(z')
\right\}\Theta(z-z')/W
$, 
where 
$\psi_-\approx (|z|+z_0)^{5/2}$ is the other independent solution for $m^2=0$,
$W=\psi'_-\psi_+-\psi'_+\psi_-$ and $\Theta(z)$ is the step function.
Then, assuming $m^2$ to be small, $\delta\psi$ 
which satisfies
$\left[-\6_z^2+V(z)\right]\de\psi=m^2\psi_+$
can be evaluated perturbatively as 
\begin{equation}
\delta\psi(z)\approx 
\frac{m^2}{W}
 \left[\psi_+\int_0^z dz'\, \psi_- \psi_+ - \psi_-\int_0^z dz'\, \psi_+^2
 \right]
\approx m^2\hat\psi_+ z^{5/2}z_0^{-2}~,  
\end{equation}
where
we used the fact that the second term in the square brackets
dominates. Without the contribution of $\delta\psi$, the solution 
turns to increase for $(|z|+z_0)/L> \pi/2$, 
and it become unnormalizable.  
In order to avoid this,  
$\delta\psi$ should be as large
as $\psi_+$ at $|z|+z_0\approx \pi L/2$. This requires 
$m^2\hat\psi_+ L^{5/2} \ell^{-2}\approx \hat\psi_+ L^{-3/2}$, 
where we used $z_0\approx \ell$ for $L\gg \ell$. 
Thus, we finally obtain $m\sim\mathcal{O}(\ell/L^2)$. 

This ``almost zero mode'' 
with mass $m\sim\mathcal{O}(\ell/L^2)$
serves as the four-dimensional 
graviton at a small distance scale. 
The effect of a small mass in four-dimensional AdS spacetime can be
estimated by considering a massive scalar field with a static source 
placed at the origin, which satisfies the equation 
$[r^{-2}\partial_r(1+r^2/L^2)r^2\partial_r-m^2]\phi=\delta^3(x)$. 
The solution with $m^2=0$ will be given by $\phi_0\approx A/r^3$ for
$r\gg L$, 
where the constant amplitude $A$ is irrelevant for our current discussion. 
Solving the first order correction due to 
the small mass iteratively, we
easily obtain $\phi_1\approx m^2 L^2 A\log r/r^3$. Comparing $\phi_0$
and $\phi_1$, 
we find that the mass effect becomes important only when $\log r
\gtrsim 1/m^2 L^2$. Since the proper radial distance in AdS is given by
$\approx L\log r$, we conclude that the mass effect 
appears at the physical length scale longer than $L^3/\ell^2$, which is much larger 
than the simple inverse mass scale 
$m^{-1}\sim\mathcal{O}(L^2/\ell)$~\cite{insidestory}. 
On the scale beyond $L^3/\ell^2$, the nature of gravity completely
deviates from the four-dimensional one.

In the KR model, the bulk opens up to the AdS boundary and its
volume diverges, if we do not put a second brane 
to cut off the bulk.
Thus, we cannot derive the effective four-dimensional Planck mass as we did 
in Eq.~(\ref{Mplrelation}). 
Reference \citen{insidestory} used shockwave solutions in the KR model to read it
out from spacetime geometry, 
and argued that the effective 
Planck mass runs as 
$M_\text{4D}^2\propto 1+\mathcal{O}(1)\left(\frac{\ell}{L}\right)^2
\frac{\mathcal{R}}{L}$\,,
where $\mathcal{R}$ is the scale of interest.
This agrees with the above observation that 
the non-zero mass of the ``almost zero mode''
starts to influence on the scale $\approx L^3/\ell^2$.

By increasing the brane tension from the RS value, not decreasing as we did 
above, we obtain the RS model with asymptotically de Sitter brane. 
For such a model, the unperturbed metric is given by Eq.~(\ref{KRmetric}) with
$L\to iL$.
As we can see from this metric, 
the spatial topology of the brane is $S^2$ and the bulk volume is finite.
Then, the ordinary four-dimensional gravity is recovered thanks to the zero
mode, and correction emerges at high energy scale due to 
massive KK modes~\cite{Gen:2000nu}.
Such braneworld models de Sitter brane
were firstly considered by Refs.~\citen{Kaloper:1999sm} and \citen{Nihei:1999mt}, 
and have been studied by many authors 
(see e.g.~Refs.~\citen{Gen:2000nu} or \citen{Iwashita:2006zj}).

\begin{figure}[t]
\centering
\subfigure[RS model with de Sitter brane%
]{%
\includegraphics[width=4cm,clip]{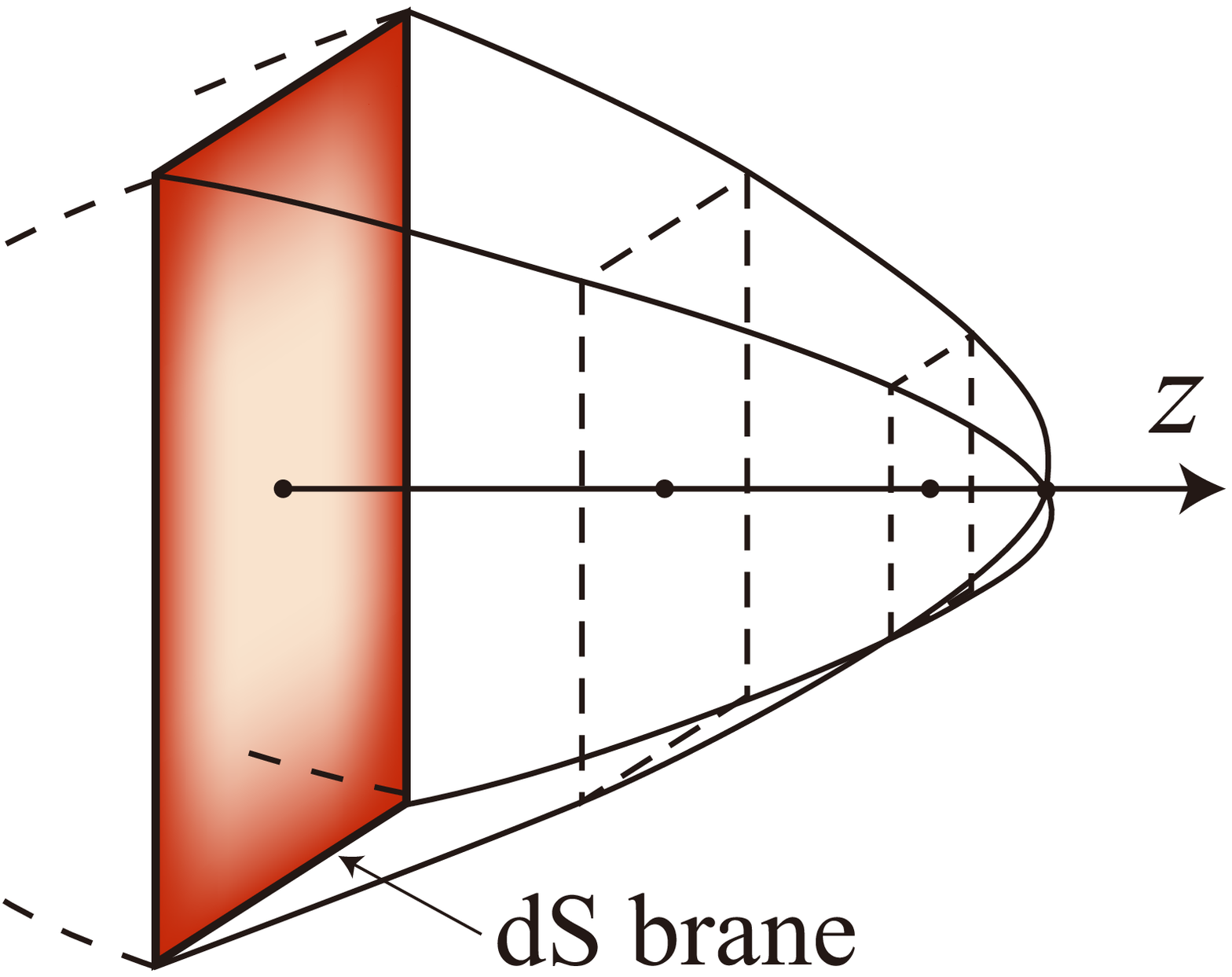}
\label{Fig:dSinGN}
}
\subfigure[RS-II model%
]{%
\includegraphics[width=4.5cm,clip]{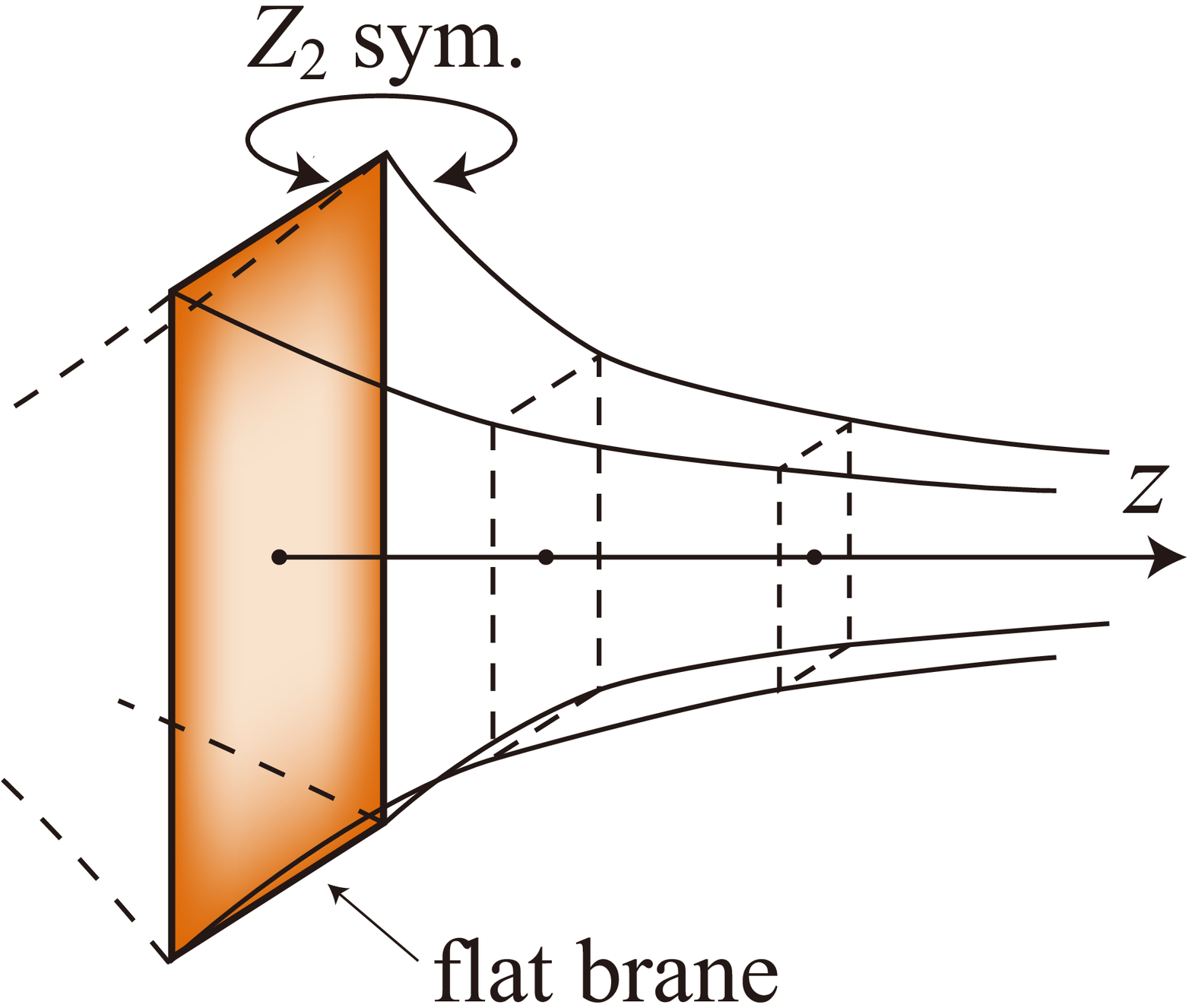}
\label{Fig:RS-IIinGN}
}
\subfigure[KR model
]{%
\includegraphics[width=4.7cm,clip]{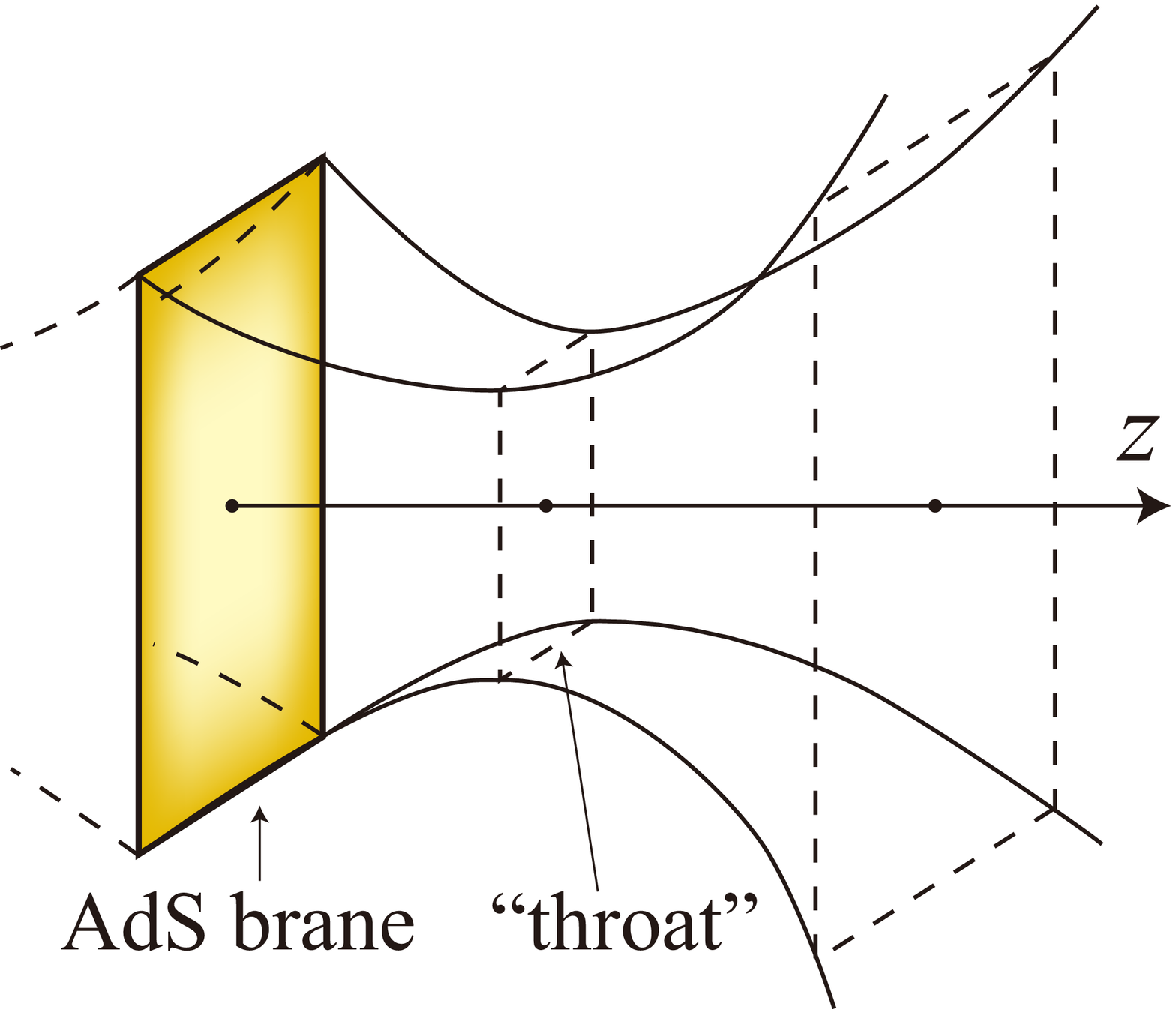}
\label{Fig:KRinGN}
}
\caption{%
Schematic of 
(a) the RS model with de Sitter brane, 
(b) the RS-II model and (c) the KR model in the \poincare coordinates.
}
\end{figure}

\subsection{DGP model}
\label{Sec:DGPbasic}

The DGP model provides another way to
localize four-dimensional graviton onto the brane~\cite{DGP}. 
The brane in this model has
its own four-dimensional Einstein-Hilbert action, which may be induced
by e.g.~quantum correction of matter fields on the brane~\cite{BD}.
The original DGP model is composed of four-dimensional brane in 
the bulk without a higher-dimensional 
cosmological constant, and its full action is given by
\begin{equation}
 S_\text{DGP} = 
-\frac{1}{16\pi G_5} \int d^5 x \sqrt{-g}\, R_\text{5D} 
- \int d^4 x \sqrt{-\ga}
\left(\frac{1}{16\pi G_4} R_\text{4D} + L_\text{matter}  \right)~.
\end{equation}
The unperturbed background is simply given by five-dimensional Minkowski 
space bounded on one side by a four-dimensional $Z_2$-symmetric
Minkowski brane.  

The basic feature of gravity on the brane in this model can 
be understood by investigating an analogous five-dimensional 
scalar field toy model whose equation of motion is given by 
\begin{equation}
 \left[G_5^{-1}\Box +\delta(y)G_4^{-1}\Box^{(4)}\right]\phi=\delta(y) J,
\end{equation}
where $J$ is the source term localized on the brane. 
$y$ is the coordinate in the direction of the extra dimension, 
and the brane is located at $y=0$. 
Using the Fourier components with respect to 
four-dimensional coordinates, which are 
expressed by associating a tilde ``$~\tilde{}~$'', 
the equation becomes 
$
 G_5^{-1} (p^2-\partial_y^2)\tilde\phi +\delta(y)G_4^{-1} p^2\tilde
  \phi=-\delta(y)\tilde J
$, where $p$ is the four-dimensional momentum parallel to the brane. 
Integrating the above equation over an infinitesimal distance across the
 brane, we obtain 
$
 -2G_5^{-1}\partial_y \tilde\phi +G_4^{-1} p^2\tilde
  \phi=-\tilde J. 
$
The solution in the bulk that is normalizable is given by 
$\tilde\phi=\tilde\phi_0 e^{-p|y|}$. Substituting this into the 
above equation, we obtain 
$
 \left[2G_5^{-1}p +G_4^{-1} p^2 \right] \tilde \phi=-\tilde J. 
$
Hence, the propagator behaves like
\begin{equation}
\tilde G(p)\sim \frac{1}{p^2 + r_c^{-1} p}~,
\label{DGPpropagator}
\end{equation}
where $r_c\equiv G_5 / 2 G_4$ defines the crossover scale for this
propagator: the propagator reduces to an
ordinary four-dimensional one in a short distance scale $p^{-1}\ll r_c$ 
while it becomes a
five-dimensional one in a long distance scale $p^{-1}\gg r_c$.

The graviton propagator in this model has a structure similar to 
Eq.~(\ref{DGPpropagator}), and then
the behavior of gravity is modified in a long distance scale while the
four-dimensional gravity is kept intact in a short distance scale in this model.
This peculiar property of the model brings about many interesting phenomena such as 
the self-accelerating cosmology~\cite{Deffayet:2000uy,Deffayet:2001pu,Sahni:2002dx}, 
and further motivated some other cosmological models~\cite{galileon}.
Many studies on the DGP cosmology showed that the self-accelerating
branch of solutions suffers from the presence of a ghost 
excitation~\cite{Koyama2005,Charmousis:2006pn,Gorbunov:2006,Izumi:2006ca,Gregory:2007xy,Gregory:2008bf}.
Here a ghost means a degree of freedom whose kinetic term has an
opposite sign, and it leads to pair production instability in quantum theory. 
However, there are also some arguments that this ghost might be less 
harmful compared with the other ghosts~\cite{Izumi:2009a,Izumi:2009b}. 

The DGP model is closely related to massive gravity 
theories~\cite{DGP,Deffayet:2001uk,Luty,Dvali:2006su,Tanaka:2010zzc}.
At the linear level, 
a four-dimensional ghost-free massive gravity theory can be defined by 
\begin{equation}
 S_m = \frac{1}{16\pi G_4} 
\int d^4 x \sqrt{-g} \biggl[
R + \frac{m_g^2}{4}\left(
h^{\m\n} h_{\m\n} - h^2
 \right)
\biggr]~,
\label{Eq:massivegravity}
\end{equation}
where $m_g$ is the graviton mass and
$h_{\m\n}\equiv g_{\m\n} - g^{(0)}_{\m\n}$ is the metric perturbation from
the background metric $g^{(0)}_{\m\n}$.
The quadratic effective action in the DGP model has a structure similar to this
one with $m_g \sim \sqrt{p/r_c}$, and thus it shares many properties with a massive
gravity theory. 
In this regard, the DGP model may be viewed as a covariant non-linear
completion of a massive gravity theory.

In the massive gravity theory defined by Eq.~(\ref{Eq:massivegravity}), 
non-zero graviton mass alters the gravitational attraction to be 
a short-ranged one. 
A perturbative analysis shows that the Newton potential in this theory is given
by $V\propto e^{-m_g r}/r$.
More striking effect of the graviton mass in linear perturbation 
is the discontinuity in the spatial component of metric perturbation from Minkowski
background,  which is so-called van Dam-Veltman-Zakharov 
(vDVZ) discontinuity~\cite{Iwasaki:1971uz,vanDam:1970vg,Zakharov:1970cc}.
Because of this discontinuity, 
even if we take the massless limit, the light bending around a star 
is predicted to be 25\% less compared with the Einstein gravity. 
The perturbative analysis, however, 
is known to break down on a short distance scale~\cite{Vainshtein}.
By solving the Einstein equation to sub-leading
order, one finds that next-to-leading order correction with respect to
$r_g$ is given by $\mathcal{O}\left( r_g /m_g^{4} r^{5}\right)$,
where $r_g$ is the Schwarzschild radius of the matter source.
Thus, the perturbative expansion with respect to $r_g$ breaks down for 
$r\lesssim r_*\equiv \left(r_g /m_g^4\right)^{1/5}$.
This critical length scale $r_*$ is called the Vainshtein scale. 
The perturbative solution will be correct for $r\gg r_*$, but 
the non-linearity plays an important role on short distance 
scales $r\lesssim r_*$. 
The vDVZ discontinuity is thought to be resolved 
by the effect of non-linearity,
which is called Vainshtein mechanism.

In the DGP model, the graviton mass is roughly given by $m_g\sim \sqrt{1/r_c r}$,
and then the Vainshtein scale becomes $r_* \sim \left(r_g r_c^2\right)^{1/3}$. 
Note that the hierarchy of the length scales becomes 
$r_g\ll r_*\ll r_c$ for an ordinary astrophysical object.
The gravitational field for $r\gg r_*$ is well approximated by a linear theory, 
and non-linearity must be taken into account for $r\lesssim r_*$.
In Refs.~\citen{Gruzinov} and \citen{TanakaDGP}, 
systematic perturbative expansion which has smooth massless limit $m_g \to 0$ was 
developed.
As a result, it was found to be important to solve the Einstein equations 
partly up to second order on a short distance scale $r\lesssim r_*$.


\section{Holographic interpretation of the warped compactification}
\label{Sec:Hol}
\subsection{Effective Einstein equations on the brane}
\label{Sec:SMS}

Before we discuss the holographic view of RS-II and KR models 
based on the AdS/CFT correspondence, 
let us study the effective  
four-dimensional Einstein equations on the brane in these models~\cite{SMS}.
As we briefly mentioned in \S\ref{Sec:RSbasic}, a brane
is characterized by its unit normal $s_M$
and the induced metric defined as $\ga_{MN}\equiv g_{MN}-s_M s_N$.
We take the convention that the normal $s^M$ points in the direction 
toward the bulk spacetime from the brane.
Another quantity that specifies the geometry of the brane is the extrinsic 
curvature, defined by $K_{\m\n}\equiv -\ga_\m^{~M}\ga_\n^{~N}\na_M s_N$,
which describes how the brane is embedded in the bulk spacetime.

The effective Einstein equations on the brane is obtained by projecting the 
five-dimensional equations onto the brane and to the normal.
The key equations for this reduction are the Gauss-Codazzi formula 
and other formulae for the projection,
%
\begin{align}
\pj{\m}{M}\pj{\n}{N}\pj{P}{\ro}\pj{\la}{Q}R^P_{~MNQ}
&=
\DDR{}^\ro_{~\la\m\n}-K^\ro_{~\m}K_{\n\la}+ K^\ro_{~\n}K_{\m\la}~,
\label{braneGauss}
\\
s^M \pj{\m}{N}R_{MN}&=D_\m K - D_\n K^\n_{~\m}~,
\label{braneCodazzi}
\\
\ga_{\m M}s^P \pj{\n}{N}s^Q R^M_{~PNQ}
&=\Lie_s K_{\m\n} + K_{\m\ro}K^\ro_{~\n}~,
\label{braneMixed}
\end{align}
where  $\DDR_{\m\n}$ and $D_\m$
are the four-dimensional Ricci tensor and the covariant derivative 
with respect to $\ga_{\m\n}$,
and $\Lie_s$ is the Lie derivative with respect to $s_M$. 
Projecting the defining equation of the Weyl tensor
\begin{equation}
C_{MPNQ}
\equiv
R_{MPNQ} 
-\frac{2}{3}\left(
g_{M[N}R_{Q]P}
-g_{P[N}R_{Q]M}
\right)
+\frac{1}{6}g_{M[N}g_{Q]P}R~, 
\end{equation}
into the directions parallel to the brane, and contracting two indices, 
one can derive an expression 
for the projected five-dimensional Einstein tensor with the aid of 
Eq.~(\ref{braneGauss}) as 
\begin{align}
\DDR_{\m\n}-\frac{1}{2}\DDR \ga_{\m\n}
&=
\frac{16\pi G_5 }{3}
\left\{
\pj{\m}{M}\pj{\n}{N}\T_{MN} + \left(
s^M s^N \T_{MN} -\frac{1}{4}\T^M_{~M}
\right)\ga_{\m\n}
\right\}
\no
&\phantom{=}\;
+KK_{\m\n}- K_\m^{~\ro}K_{\n\ro}
-\frac{1}{2}\left(
K^2-K^{\ro\la}K_{\ro\la}
\right)\ga_{\m\n}
-E_{\m\n}~,
\label{projectedG}
\end{align}
where we used the five-dimensional Einstein equation 
\begin{equation}
R_{MN}-\frac{1}{2}R g_{MN}=8\pi G_5 T_{MN}~,
\qquad
\T_{MN} \equiv 
\frac{-\Lambda_\text{5D}}{8\pi G_5}g_{MN}+ \left(
-\la \ga_{MN}+ \tau_{MN}
\right)\de(y)~,
\end{equation} 
in which $\la$ and $\tau_{MN}$ are the brane tension and the energy-momentum 
tensor confined on the brane, as before.
The coordinate $y$ in the delta function here is that of the Gaussian-normal 
coordinates with respect to the brane, in which the five-dimensional metric is 
expressed as 
\begin{equation}
ds^2=dy^2+\ga_{\m\n}dx^\m dx^\n~.
\label{genuineGN}
\end{equation}
(Here, for notational simplicity, we do not put a bar
``$~\bar{}~$'' to the quantities in the Gaussian-normal coordinates.) 
A projection of the five-dimensional Weyl tensor 
$E_{\m\n}$ in Eq.~(\ref{projectedG}) is 
defined by  
\begin{align}
E_{\m\n} &\equiv
C^M_{~NPQ} \,s_M  \pj{\m}{N} s^P \pj{\n}{Q}~.
\end{align}

To derive the junction condition at the brane, let us rewrite 
the left-hand side of
Eq.~(\ref{projectedG}) using Eqs.~(\ref{braneGauss}) and (\ref{braneMixed}) as 
\begin{align}
\DDR_{\m\n}-\frac{1}{2}\DDR\ga_{\m\n}
&= \pj{\m}{M}\pj{\n}{N}\left(
R_{MN}-\frac{1}{2}Rg_{MN}
\right)
+ KK_{\m\n}-K_{\m\ro}K^\ro_{~\n}-\Lie_s K_{\m\n}
\no
&\phantom{=}\;
+ \frac{1}{2}\left(
-K^2 + 2\Lie_s K -K_{\ro\la}K^{\ro\la}
\right)\ga_{\m\n}~.
\label{preJC}
\end{align}
In the Gaussian-normal coordinates~(\ref{genuineGN}), the Lie derivative $\Lie_n$ 
reduces to a simple coordinate derivative $\6_y$. 
The terms other than this Lie derivative and the delta function 
concealed in the five-dimensional Einstein tensor are all 
regular on the brane. Then, integrating Eq.~(\ref{preJC})
along $y$ over an infinitesimal domain $y\in(-\epsilon,\epsilon)$, we
recover Eq.~(\ref{RS-IJC}). 
This equation is called the Junction conditions for the bulk metric~\cite{Israel:1966rt}.

Since $Z_2$-symmetry across the brane demands 
$\lim_{y\to+0}K_{\m\n}=-\lim_{y\to-0}K_{\m\n}$, we find that the junction 
conditions~(\ref{RS-IJC}) reduce to 
\begin{equation}
K_{\m\n}-K\ga_{\m\n}=4\pi G_5
\left(
-\la\ga_{\m\n} + \tau_{\m\n}
\right)~,
\label{genuineJC2}
\end{equation}
where $K_{\m\n}$ is understood as $\lim_{y\to+0}K_{\m\n}$. 
Using this formula, we can transform Eq.~(\ref{projectedG})
further  into
\begin{equation}
\DDR_{\m\n}-\frac{1}{2}\DDR\ga_{\m\n} = -\La_\text{4D}\ga_{\m\n}
+ 8\pi G_4 \tau_{\m\n} + \left(8\pi G_5\right)^2 \pi_{\m\n} -E_{\m\n}~,
\label{effectiveEeq}
\end{equation}
where the four-dimensional cosmological constant and the Newton constant are 
given by
\begin{equation}
\La_\text{4D}=\frac{1}{2}\left(
\La_\text{5D}+ \frac{\left(8\pi G_5\right)^4}{6}\la^2
\right)~, 
\qquad
G_4 = \frac{4\pi G_5\la}{3}~,
\end{equation}
and 
\begin{equation}
\pi_{\m\n}\equiv -\frac{1}{4}\tau_{\m\ro}\tau_{~\n}^\ro 
+ \frac{1}{12} \tau \tau_{\m\n}
+ \frac{1}{8} \tau_{\ro\la}\tau^{\ro\la}\ga_{\m\n} 
- \frac{1}{24} \tau^2 \ga_{\m\n}
\label{piTerm}
\end{equation}
is the contribution quadratic in $\tau_{\m\n}$. 
As we can see in Eq.~(\ref{effectiveEeq}), 
a brane observer sees an 
effective energy-momentum tensor $\left(6 G_5\right/\lambda)
\pi_{\m\n} 
-(8\pi G_4)^{-1}E_{\m\n}$.
Since $E_{\m\n}$ is traceless by its construction, and 
the trace part coming from $\pi_{\m\n}$ term can be identified 
with the trace anomaly~\cite{Shiromizu:2001ve}, 
this effective energy-momentum tensor may be regarded as 
the contribution from some conformal fields, which we will discuss in the next section.
References \citen{Kanno:2002iaa} and \citen{McFadden:2004se}
constructed 
the effective action on the brane that derives Eq.~(\ref{effectiveEeq}),
and argued their relationship with the AdS/CFT correspondence
(for a review see Ref.~\citen{Soda:2010si}).

The effective Einstein equations on the brane in the DGP model was studied in
Ref.~\citen{MizunoDGP} in detail. 
The DGP model can be seen as a variant of the RS model 
with exotic matter whose Lagrangian is
the four-dimensional Ricci scalar. Then, the modification to the effective
Einstein equation only appears in the brane energy-momentum tensor.
For the DGP model, whose brane-localized part of the Lagrangian is 
\begin{equation}
 L_\text{brane} = \frac{1}{16\pi G_4} R_\text{4D} - \lambda + L_\text{matter}~,
\end{equation}
the energy-momentum tensor on the brane is given by
\begin{equation}
 \tau_{\m\n} = -2 \frac{\de L_\text{brane}}{\de \ga^{\m\n}} + \ga_{\m\n}
  L_\text{brane}
= - \la \ga_{\m\n} + T^\text{matter}_{\m\n} - \frac{1}{8\pi G_4}G^\text{4D}_{\m\n}~.
\end{equation}
We will immediately obtain the effective Einstein equations for the DGP model by
plugging this $\tau_{\m\n}$ into Eq.~(\ref{effectiveEeq}) with the 
bulk cosmological constant being set to zero.
In Ref.~\citen{MizunoDGP}, this effective equations were used to
study various settings in the DGP model.

\subsection{AdS/CFT correspondence in the RS-II braneworld model}
\label{Sec:AdSCFTinRS}

In the preceding sections, we introduced the braneworld models that have 
the AdS spacetime as the unperturbed bulk spacetime, and surveyed 
basic properties of gravity in these models.
In this subsection, we focus on another aspect of these models, which is the 
non-trivial correspondence between the classical gravity in higher
dimensions and the gauge theory on the brane 
suggested from the AdS/CFT correspondence.

In the RS-II model,
the brane serves as a boundary of the asymptotically AdS bulk.
Then, naively thinking, it might be possible to apply to this model 
the AdS/CFT correspondence, which implies the correspondence between the 
classical supergravity and the conformal field theory on the AdS 
boundary~\cite{Mal,Aharony:1999ti,Maldacena:2003nj,Freedman,Johnson,Becker}.
Such a proposition was firstly made by Refs.~\citen{WittenComment}
and \citen{VerlindeComment}, and 
made more concrete in Refs.~\citen{Gubser,HR,Savonije:2001nd}.

We review below the discussion of Ref.~\citen{HR}
about the relationship between the RS model and the CFT on the brane.
The AdS/CFT correspondence states that the supergravity partition function,
$Z[\bfga]$, is related to the generating functional of connected Green's 
functions for the CFT on the boundary geometry, 
$W_\text{CFT}[\bfga]$, as 
\begin{equation}
Z[\bfga]= \int d[\bfg]\exp\left(
i S_\text{grav}[\bfg]
\right)
= 
\int d[\phi]\exp\left(
iS_\text{CFT}[\phi;\bfga]
\right)
\equiv
\exp\left(
iW_\text{CFT}[\bfga]
\right)~,
\label{thecorrespondence}
\end{equation}
where $\ga_{\m\n}$ is the conformally-rescaled boundary metric and $\phi$ 
symbolically represents the fields contained in the CFT. 
The supergravity action for the bulk spacetime, $S_\text{grav}$, is given by
\begin{equation}
S_\text{grav}=S_\text{EH} + S_{GH} + S_1 + S_2 + S_3~,
\label{Sgrav}
\end{equation}
where 
\begin{equation}
S_\text{EH} = \frac{1}{16\pi G_5} \int d^5 x \sqrt{-g}\left(R+2\La\right)~,
\quad
S_\text{GH} = \frac{1}{8\pi G_5}\int d^4x \sqrt{-\ga}K
\end{equation}
are respectively 
the Einstein-Hilbert action for the bulk gravity
and the Gibbons-Hawking boundary term 
introduced to make the variational problem to be well-defined~\cite{GH}.
When we move the boundary toward the AdS boundary, 
this $S_\text{EH}$ diverges since the volume of an AdS spacetime is infinite.
To regularize it, the counter terms, $S_1, S_2$ and $S_3$, are introduced to
the gravitational action~(\ref{Sgrav}).
To be more precise, 
let us consider the asymptotic AdS metric 
\begin{equation}
 ds^2={\ell^2\over z^2}\left(\gamma_{\mu\nu}dx^\mu dx^\nu +dz^2\right)~. 
\end{equation}
First, we cut 
the region with $z<\epsilon\ll \ell$ from this spacetime. 
Next, we expand the gravitational action in terms of the small parameter
$\epsilon$.
Then, we obtain some terms that diverge in the limit of $\epsilon\to 0$.
The negatives of such terms are given by 
\begin{gather}
S_1 = \frac{-3}{8\pi G_5 \ell}\int d^4x\sqrt{-\ga}~,
\quad
S_2 = \frac{-\ell}{32\pi G_5}\int d^4x\sqrt{-\ga}\DDR~,
\no
S_3 = \frac{-\ell^3\log\epsilon}{64\pi G_5}\int d^4x \sqrt{-\ga}\left(
\DDR_{\m\n}\DDR{}^{\m\n}-\frac{1}{3}\DDR{}^2
\right)~, 
\end{gather}
and are identified with the counter terms. 

Now, let us consider the RS-II model, whose action is in the 
current notation given by
\begin{equation}
S_\text{RS} = S_\text{EH} + S_\text{GH} + 2S_1 + S_m~,
\end{equation}
where $S_m$ is 
the action of the matter confined on the brane. 
$2S_1$ is precisely corresponding to the action for the brane tension $3/4\pi
G_5 \ell$. 
Let us construct the partition function from this $S_\text{RS}$. It is given as 
\begin{align}
Z[\bfga] &= 
\int d[\de \bfg]d[\phi]\exp\left(iS_\text{RS}[\bfg_0+\de\bfg]\right)
\no
&= \exp(2iS_1[\bfga_0+\bfga ])
\no
&\phantom{=}\;\;
\times
\int  d[\de \bfg]d[\phi] \exp\left(
iS_\text{EH}[\bfg_0+\de\bfg] + iS_\text{GH}[\bfg_0+\de\bfg] 
+ iS_m[\phi;\bfga_0+\bfga]
\right)
\no
&=
\exp(2iS_1[\bfga_0+\bfga ])
\left(
\int_\text{\rlap{one-bulk}}
\;
  d[\de \bfg]d[\phi] \exp\left(
iS_\text{EH}[\bfg_0+\de\bfg] + iS_\text{GH}[\bfg_0+\de\bfg] 
\right)
\right)^2
\no
&\phantom{=}\;\;
\times 
\int d[\phi]\exp\left(
iS_m[\phi;\bfga_0 + \bfga]
\right)~,
\label{Zpre}
\end{align}
where in the third equality we used 
the fact that the RS-II model has two bulk regions which are 
related by $Z_2$-symmetry with each other.
Now, we apply Eq.~(\ref{thecorrespondence}) of the AdS/CFT correspondence,
which holds for the regularized gravitational action~(\ref{Sgrav}), to this
equation. Firstly, Eqs.~(\ref{thecorrespondence}) and (\ref{Sgrav}) imply
\begin{align}
\int_\text{\rlap{one-bulk}}~
  d[\de \bfg]d[\phi] \exp\left(
iS_\text{EH} + iS_\text{GH}
\right)[\bfg_0+\de\bfg]
&=
\exp\left(
iW_\text{CFT}-iS_1-iS_2-iS_3
\right)[\bfga_0 + \de\bfga]~.
\end{align}
Plugging this into Eq.~(\ref{Zpre}), we obtain 
\begin{align}
Z[\bfga]= \exp\left(
2iW_\text{CFT} -2iS_2-2iS_3
\right)[\bfga_0 + \de\bfga]
\int d[\phi]\exp\left(
iS_m[\phi;\bfga_0 + \bfga]
\right)~.
\label{Z}
\end{align}
Remarkably, $-2S_2$ is equal to the four-dimensional Einstein-Hilbert action.
Thus, the system described by the partition function on the right-hand side is 
that of 
the four-dimensional CFT interacting with the four-dimensional gravity and 
the matter described by $-2S_2$ and $S_m$, respectively. 
The other term, $S_3$, is 
the higher-curvature correction to the theory, and it represents 
the renormalization scale dependence of the cutoff CFT.

This correspondence in the RS-II model is an extension of the original AdS/CFT
correspondence in the sense that 
the former reduces to the latter in the limit of $\epsilon\to 0$. 
Namely, 
the correspondence in the RS-II model is different from the original one
in that the boundary is not sent to the AdS boundary 
but is determined by the brane position.
The position of the boundary in the bulk corresponds to 
the UV cutoff energy scale of the CFT. 
When we measure the energy scale by the proper length on the brane, where the warp
factor is usually normalized to unity, the UV cutoff energy scale is $\sim
\ell^{-1}$. If we send the brane to the AdS boundary, the cut-off energy
scale becomes higher. However, this shift of the location of the brane 
can be absorbed by the overall re-normalization of the warp factor. 
Therefore, sending the boundary to the AdS boundary is 
equivalent to observe phenomena in the IR limit. 
Hence, we expect that this extended correspondence should hold at 
least for deep IR phenomena compared with the bulk curvature scale.

\subsection{Supporting evidence for the correspondence}
\label{Sec:StudiesOnAdSCFTinRS}

Although there is no proof of this extended correspondence in the 
RS-II model, there are
many pieces of supporting evidence.
We show some examples of them below.

\begin{itemize}
\item Garriga and Sasaki\cite{Garriga:2000} considered a bulk black hole 
in the expanding braneworld. The presence of a bulk black hole 
naturally induces additional thermal radiation of CFT to the boundary 
effective theory. When the temperature is well below the AdS scale, 
the entropy of this radiation agrees with the entropy determined from 
the area of the black hole in the AdS bulk. 

\item Duff and Liu~\cite{Duff:2000mt} calculated weakly coupled 
CFT's one-loop correction to the graviton propagator, 
and found that the Newton potential acquires a correction suppressed 
by a factor of $\mathcal{O}(\ell^2/r^{2})$.
This correction including its coefficient agrees with 
that in the RS-II model, Eq.~(\ref{NewtonPotential}),
which is caused by five-dimensional gravity in the bulk~\cite{Garriga:1999yh}.
This coincidence between four and five-dimensional description was reconfirmed
by Anderson et al.\cite{Anderson:2004md,Fabbri2006}~using 
vacuum polarization of various fields
around a Schwarzschild black hole.
%

\item Shiromizu et al.\cite{Shiromizu:2001ve}~studied the relationship between 
CFT energy-momentum tensor 
in the four-dimensional point of view and 
effective energy-momentum tensor on the brane
in the five-dimensional point of view.
As a result, 
the CFT's trace anomaly 
was found to be reproduced 
in the five-dimensional point of view
from the $\pi_{\m\n}$ term in the effective Einstein equations,
Eq.~(\ref{piTerm}),
as $\left<T^\m_{~\m}\right>_\text{CFT}=\ell \pi^\m_{~\m}/4$.

\item Tanaka\cite{TanakaAdS} studied the Friedmann-Lemaitre-Robertson-Walker 
cosmology realized on the braneworld model.
It was shown 
that the effective equations of motion for four-dimensional
tensor perturbation on the brane, which is obtained in the five-dimensional classical 
gravity picture, agree with those derived in the four-dimensional CFT picture.

\item Grisa and Pujol\`as\cite{Oriol} constructed solutions of 
three-dimensional domain wall localized on the 
four-dimensional brane in the RS-II model 
to use it as a testbed of the correspondence.
In the four-dimensional point of view, the domain wall tension is renormalized 
      due to the CFT's trace anomaly.
They found that
this renormalization of tension is reproduced
in the five-dimensional point 
of view as an effect of brane acceleration in the five-dimensional bulk.
\end{itemize}

We also have interesting observations, although they are not the 
evidence directly supporting the argument that five-dimensional 
gravity phenomena is described by four-dimensional general relativity 
with CFT corrections. 
Let us consider four-dimensional de Sitter spacetime.
This de Sitter spacetime is associated with finite temperature due to existence of
cosmological horizon. 
Das et al.\cite{Das:2001zc}~considered the RS model with de Sitter brane, and argued
that this de Sitter temperature on the brane
can be reproduced in the five-dimensional
point of view from accelerating motion of the brane in the bulk.
Iwasita et al.\cite{Iwashita:2006zj}~also studied this model, and 
showed that the holographic 
entanglement entropy in the five-dimensional picture
coincides with 
the four-dimensional de Sitter entropy calculated from the 
cosmological horizon area or from the Euclidean path integral.

Assuming the correspondence to be correct, some attempts to give predictions 
about the cosmology realized on the RS model were made.
Reference \citen{HR} applied the correspondence to inflating universe and quantum
fluctuation of strongly-coupled large $N$ CFT therein.
They constructed an effective action of the CFT applying the
correspondence to the RS model with de Sitter brane.
An advantage of this approach is that the effect of full quantum effect of the
matter fields, which is usually out of reach of perturbative approaches, can be
taken into account.
From the effective action, they showed that the tensor 
perturbation due to the quantum effect is highly suppressed in short
scale.
Since any matter field can be expected to behave like CFT in high energy scale, 
such a prediction may apply to general matter fields which are strongly coupled in
such regime.

An upper bound on $\ell$ was given from the production of CFT particles 
in Ref.~\citen{Gubser}
assuming the correspondence to hold,
and 
this bound was re-analyzed later by Ref.~\citen{Hebecker:2001nv} 
more explicitly based on the five-dimensional gravity picture.
They considered scattering of standard model particles and CFT in 
the Friedmann universe.
In the five-dimensional gravity picture, the corresponding 
process is the bulk graviton emission from the 
radiation on the Freedman-Robertson-Walker brane. 
When the center-of-mass energy of the standard model particles 
$\sqrt{s}$ is much higher than the AdS bulk energy scale $1/\ell$, 
those particles will feel the five-dimensional gravity. 
Thus, its cross section to emit the bulk graviton will be
proportional to $G_5$, and we may expect 
the cross section to be
$\s(s)\sim G_5 \sqrt{s}$ from the dimensional analysis.
Multiplying the effective number of the brane matter's degrees of freedom, 
we obtain the total cross section.
From this cross section,
taking into account the redshift factor properly,  
one can estimate the energy leakage to the bulk graviton.
This bulk graviton will be seen by a brane observer as 
``dark radiation'' which interacts only 
gravitationally with other particles.
The dark radiation density $\ro_d$, which turns out to 
depend logarithmically on the maximum temperature of the universe,  
has to be smaller than the bound from the nucleosynthesis. 
This bound marginally accepts $\ell\approx 1\,\text{mm}$ even 
if the maximum temperature of the universe 
is as high as the fundamental energy scale $M_5$. 
They showed that this energy leakage 
into the bulk graviton can be equivalently viewed as 
the leakage into the large $N$ CFT,
and also pointed out that 
cosmological implications at high energy scale may depend on microscopic
definition of the correspondence, i.e., the background spacetime, 
such as AdS$_5$ or AdS$_5\times S^5$, on which we consider the model
(see also Ref.~\citen{ArkaniHamed:2000ds}).

\subsection{Holography in the KR model}

Since the KR model can be obtained as a continuous deformation of the RS-II 
model, it is natural to expect that the bulk/brane correspondence 
 persists even in the KR model.
Such a view was already taken in the original paper of the KR 
model~\cite{KR} and has been investigated in many ways.
For example, 
the ``almost zero mode'' 
in the KR model
we mentioned in \S\ref{Sec:KRbasic}
was reproduced in four-dimensional point of view as a
graviton dressed with
one-loop correction of the CFT~\cite{P,Porrati:2003sa,Porrati:2001gx,Duff}.
This issue is also pursued in Ref.~\citen{insidestory}, and they interpreted 
the running of the effective four-dimensional Planck mass,
$M_\text{4D}^2\propto 1+\mathcal{O}(1)\left(\frac{\ell}{L}\right)^2\frac{\mathcal{R}}{L}$,
as renormalization group flow driven by quantum effects of the CFT.
In Refs.~\citen{KRstring_orig,KRstring,DeWolfe,DeWolfe2}, 
the realization of the KR 
model within the string theory and the AdS/CFT correspondence in such a system 
with defects on the AdS boundary are discussed.

\begin{wrapfigure}{r}{6.6cm}
\centering
\includegraphics[width=5.6cm,clip]{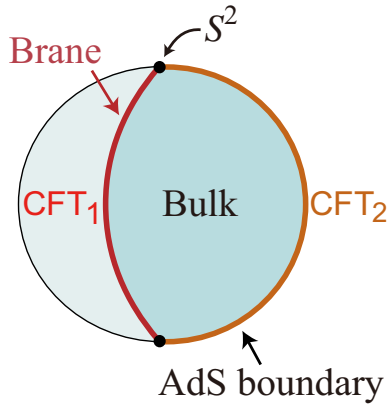}
\caption{%
A schematic picture of the a conformally transformed 
section of the KR model sliced at a time constant surface. 
The brane touches the AdS boundary, forming a $S^2$ boundary between them.
CFT$_1$ is living on the brane while CFT$_2$ on the AdS boundary. 
}
\label{Fig:KRbrane}
\end{wrapfigure}
In the KR model with a single brane, 
there are two boundaries surrounding the bulk: the brane and the half of the 
AdS boundary as depicted in Fig.~\ref{Fig:KRbrane}. 
Then, the bulk gravity will induce one set of CFT on each boundary and there will be
two sets of CFT in total.
Naively thinking, each CFT would have distinct bulk region which is dual to it.
In Ref.~\citen{BR}, such an idea was pursued by considering causal structure
of the bulk spacetime. As a result of this study,
it was realized that the ``holographic domains'' of the bulk, which 
are the bulk regions corresponding to the respective CFTs on the boundaries, 
are given by the bulk regions divided at the ``throat'' of the bulk at which the 
warp factor is minimized. 
Since gravitational waves can freely pass through this
throat in the five-dimensional description, it is expected in the four-dimensional
picture that the CFTs on the brane and the AdS boundary communicate each
other through the common $S^2$ boundary. 
This transparent boundary condition at the common boundary 
is found to be essential to reproduce the graviton mass by evaluating 
the correction coming from the CFT loop~\cite{P,Porrati:2003sa}.

\section{Black holes in ADD model: Kaluza-Klein black holes}
\label{Sec:ADDBH}

In this and the subsequent sections,
we discuss black hole solutions in the braneworld models.
Let us begin with the ADD model, which is the simplest braneworld model.
Since the brane in the ADD model does not have tension, the black objects 
in this model are almost equivalent to those in a Kaluza-Klein compactified spacetime. 
The properties of such black objects have been studied intensively and 
summarized in reviews such as Refs.~\citen{Kol} and \citen{Harmark_review}.
In this section, 
we only summarize some points important for discussions in the following
sections, and also we pick up some topics that is not discussed in the above
reviews. 

\subsection{Black holes escaping into the bulk}

One of the interesting predictions of braneworld models 
with its fundamental gravity scale at TeV is
that small black holes can be created in high energy collisions of
particles at energy scale within the reach of forthcoming 
experiments~\cite{r3}. When two particles on a brane 
collide at a center of mass energy larger than the fundamental gravity 
scale with sufficiently small impact parameter, 
the system will form a black hole. 
The black hole formed in this process is small compared with the 
size of extra dimensions in the context of the ADD braneworld. 
Therefore the role of the brane and the finiteness of the bulk 
is not relevant in most respects. 
Properties of such black holes smaller than extra dimension size are summarized
in references such as Ref.~\citen{Kanti:2004nr}.
Here we just mention an aspect which is peculiar to the 
braneworld setup: escaping black holes from the brane.  

After its formation, the black hole will start to emit Hawking
radiation partly into higher-dimensional bulk modes~\cite{r5}. 
The emission of higher-dimensional modes will
cause the recoil of the black hole into the extra dimensions 
unless the process is prohibited by symmetry, e.g., 
$Z_2$-symmetry in codimension-one case. 
The interaction between a small
black hole and a brane when the black hole is kicked relative to the
brane was first studied in Refs.~\citen{art1} and \citen{art1-b}. 
It was pointed out that the black hole 
can escape from the brane due to emission of higher-dimensional modes, 
using a field theory model in which such a black hole is described as a
massive scalar particle with internal degrees of freedom.  
If the black hole leaves the brane, Hawking radiation into the brane 
modes suddenly terminates and it will leave a distinctive experimental 
signature. 
In Ref.~\citen{art2} the same problem was studied 
by investigating the dynamics of Nambu-Goto branes 
in black hole spacetimes. This is to be called  
the {\it probe-brane} approximation, in the sense that 
the tension of the brane is assumed to be negligibly small. 
The results of Ref.~\citen{art2} confirmed the basic claim in 
Ref.~\citen{art1}, further suggesting that 
the escape of the black hole occurs through reconnection of the brane. 
A small piece of the brane attached to the black hole is pinched off 
from the main body of the brane. 
In Ref.~\citen{art3}, these results were confirmed also in a model in which 
the brane is described by a domain wall composed of a scalar
field. 
In Ref.~\citen{Flachi:2006ev}, the critical escape velocity 
was derived by studying the interaction between a Nambu-Goto
brane and a black hole assuming adiabatic (quasi-static)
evolution. It was found that, to lowest order effects in the tension, 
the critical escape velocity does not exist for
codimension-one branes, while it does for higher codimension branes. 
These results can be qualitatively understood by considering 
the gravitational force caused by a light brane on a flat background 
spacetime perturbatively. A codimension-one brane produces repulsive force 
while a brane with codimensions higher than 2 makes attractive force.  
Codimension-two branes are neutral in this test particle approximation 
as in the well-known case of cosmic strings in four dimensions, but 
finite amount of initial velocity is required to escape when treated 
in the opposite limit, i.e., in the test brane approximation.

\subsection{Solution sequences of KK black objects}

We consider below the black objects in a KK-compactified spacetime which is 
asymptotically $\mathcal{M}^d\times S^1$, where $\mathcal{M}^d$ is 
$d$-dimensional Minkowski spacetime.
There are typically two types of black objects in this model: a black string
wrapping the bulk and a black hole localized in the bulk.
These solutions can be parametrized by its mass 
and length along the KK direction at infinity.

Properties of localized black holes for $d=3$ was studied, e.g., in 
Refs.~\citen{Bogojevic:1990hv,Korotkin:1994dw,Frolov:2003kd}.
For $d\geq 4$, numerical solutions have been constructed 
in, e.g., Refs.~\citen{Kol:2003if,Sorkin,Kudoh:2003ki,KudohBS}.
As for analytical approaches, 
perturbative solutions were constructed by
Refs.~\citen{Gorbonos:2004uc} and \citen{Karasik:2004ds}  
using the matched asymptotic
expansion technique, and by Ref.~\citen{Harmark:2003yz} using special
coordinates which interpolate spherical coordinates near the black hole and
cylindrical coordinates at infinity~\cite{Harmark:2002tr}.

We also have a black string as another black object in KK spacetime.
In the present setup, black string solutions are given by 
\begin{equation}
 ds^2=dy^2+\ga_{\mu\nu}dx^\m dx^\n,
\end{equation} 
where $y$ is the coordinate in the direction of $S^1$ and 
$\ga_{\m\n}$ is any black hole solution of the $d$-dimensional 
vacuum Einstein equations. 
This solution suffers from instability when, roughly speaking, 
the size of 
the horizon radius becomes smaller than the length of $S^{1}$. 
This instability is called the Gregory-Laflamme 
instability~\cite{Gregory:1993vy,Gregory:1994bj}. 
A marginally unstable mode arises 
when the parameters in the solutions are tuned to the critical values.
Existence of such a marginally unstable mode indicates the 
appearance of a new branch of solutions. 
Namely, a solution perturbed in the direction of the marginally unstable mode
will also be a static solution, and the sequence of solutions will
continue to exist to the non-linear level. 
The solutions in this new branch will be non-uniform static black strings in
this case.
The properties of this non-uniform string was studied from both analytical
side~\cite{Gubser:2001ac,Sorkin:2004qq,Wiseman:2002zc}
and numerical side~\cite{Wiseman:2002zc,Kleihaus:2006ee,Sorkin:2006wp}. 
It was found that, as long as $d\leq 13$,
this non-uniform black string has larger mass
compared to a uniform black string with the same entropy,
and then it cannot be an endpoint of the Gregory-Laflamme instability in a
micro-canonical ensemble.

The first construction of numerical solutions was done 
in the following way by Wiseman~\cite{Wiseman:2002zc},
based on the numerical technique developed in Ref.~\citen{Wiseman:2001xt}. 
A static metric in axisymmetric five-dimensional spacetime can 
be expressed without loss of generality as 
\begin{equation}
 ds^2 = \frac{\ell^2}{z^2}
\Bigl(
-T^2 dt^2 + e^{2R}\left(
dr^2 + dz^2
\right) + r^2 e^{2C} d\Omega^2
\Bigr)~,
\label{conformalAnsatz}
\end{equation}
where $r$ and $z$ are the radial coordinate along the brane and 
the one in the direction of the extra dimension, respectively.
An advantage of this metric form is that the two-dimensional part spanned by $r$
and $z$ is only specified to be conformally flat, and then we may set 
the locations of both the brane and the horizon to be constant coordinate 
surfaces by applying conformal transformation.
The Einstein equations become elliptic partial differential equations for free
functions $T$, $R$ and $C$, and one can solve them with appropriate 
boundary conditions at the brane, axisymmetric axis and asymptotic boundary.
Solving them numerically by the relaxation method,
solutions for non-uniform black strings were
successfully constructed. 

\begin{wrapfigure}{l}{6.6cm}
\centering
\includegraphics[width=6.0cm,clip]{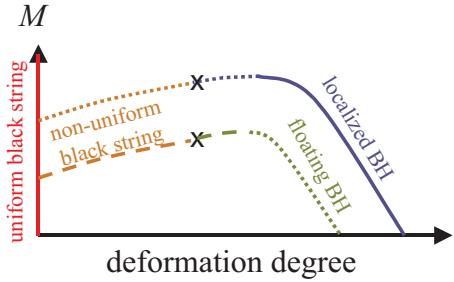}
\caption{%
Solution sequences in the braneworld models with flat bulk.
The horizontal axis denotes the deformation degree of the black object compared
 to a uniform black string.
(this figure is taken from Ref.~\citen{Tanaka:2009zz}).
}
\label{Fig:fig1_tanaka}
\end{wrapfigure}

This non-uniform black string branch is thought to be 
connected to the localized black hole
branch in the following manner. 
As we increase the deformation of a black string further, a
cone-like structure develops where the radius of the black string 
shrinks to zero. 
After this cone-like structure formation,
the sequence continues to the sequence with two isolated
segments of the horizon, and this latter sequence corresponds to
the localized black hole branch.
Such a topology-changing transition in the solution sequence was proposed
in Ref.~\citen{Kol:2002xz}.
Incorporating this topology change, the phase diagram of black objects 
in KK-compactified spacetime would be as shown in Fig.~\ref{Fig:fig1_tanaka}. 
Now numbers of studies, such as Refs.~\citen{Kudoh:2003ki} and \citen{KudohBS} 
which constructed numerical
solutions about the entire solution branch, are supporting this phase
diagram, although 
it is still difficult to resolve the transition numerically. 

Another direction of study about black string in KK spacetime is to simulate its
dynamical fate after the
Gregory-Laflamme instability.
Reference \citen{Horowitz:2001cz} showed
that it takes infinite affine time along the horizon 
for a black string to pinch off,
assuming that any singularity  
is not formed on or outside the horizon.   
In Ref.~\citen{Marolf:2005vn} it was pointed out that 
such pinching off can occur in a finite asymptotic time 
if a singularity is formed on the horizon. 
The dynamical simulation of a deforming black string is pioneered by 
Refs.~\citen{Choptuik:2003qd} and \citen{Garfinkle:2004em}, and recently a new
result was reported which suggests that a black string shrinks to 
infinitesimal thickness leaving blob-like black holes~\cite{Lehner:2010pn}.
The latter simulation also showed that this pinching process proceeds 
in a fractal manner, letting the string thinner and producing smaller 
black holes. They found that the time scale of this process is 
roughly proportional to local string radius, 
and then argued that a black string might pinch off in a finite asymptotic time.

When the extra dimension is not $S^1$ but $R^2$, the issue of finding a
brane-localized black hole solution considerably 
simplifies since codimension-two objects with tension
can be easily created by introducing deficit angle in an angular coordinate.
A black hole solution localized on a codimension-two brane 
with tension are given just by introducing deficit angle 
to the higher-dimensional Schwarzschild solution~\cite{Kaloper:2006ek}.

\section{Black holes in RS-II model}
\label{Sec:arguments}

Our direct knowledge about black hole solutions in the RS-II model 
is quite limited. Hence, 
in \S\ref{Sec:HolographicPrediction},
we first discuss the black 
hole solutions in the RS-II model applying 
the bulk/brane correspondence we introduced in \S\ref{Sec:Hol}. 
By doing so, we will obtain a prediction dubbed as the classical evaporation
conjecture of the brane-localized black hole. 
We also discuss some connection to the black objects in the ADD model,
and mention astrophysical constraints on the bulk curvature scale
given by this conjecture.
In \S\S\ref{Sec:RS-analytic} and \ref{Sec:RS-numerical},
we summarize studies on the black holes in the RS-II model and present
supporting evidence for the conjecture.
We focus on analytic approaches in \S\ref{Sec:RS-analytic} and numerical 
approaches in \S\ref{Sec:RS-numerical}.

\subsection{Holographic predictions}
\label{Sec:HolographicPrediction}

\subsubsection{Classical evaporation conjecture of brane-localized BHs}
\label{Sec:classicalEvaporation}

If the duality in the RS-II model works even in the presence of a black hole, 
an interesting implication about its time evolution will be obtained 
as follows~\cite{tanaka,emp1}.
Starting point of the argument is the properties of a four-dimensional black
hole in the brane point of view.
In the bulk/brane correspondence in the RS-II model, a large number of 
degrees of freedom of the quantum field theory 
(QFT) of 
$\mathcal{O}(\ell^2/G_4)$ 
couple to the four-dimensional gravity. Such QFT degrees of
freedom open a channel for the 
four-dimensional black hole to radiate its energy as the Hawking 
radiation very efficiently~\cite{Hawking:1974sw}. 
Then, that four-dimensional black hole will shrink and cannot remain 
static. 

Since we are assuming the bulk/brane correspondence to hold,
a five-dimensional counterpart of this shrinking four-dimensional black hole 
must exist. 
The point is 
that the bulk/brane correspondence indicates that 
this process should be described not by a quantum effect but 
by a classical dynamics in the five-dimensional picture,
and the five-dimensional counterpart should be a classical black object in 
the bulk.
Naively thinking, such a five-dimensional object will be a five-dimensional 
black hole localized on the brane, since it is a natural object to form as a 
consequence of a gravitational collapse on the brane.
If so, according to the correspondence,
the five-dimensional brane-localized black hole 
cannot be static and it will continuously reduce its area of the intersection with the 
brane as a result of classical dynamics.

We do not know much about such a dynamical 
process in five-dimensional classical gravity, 
but a warped black string solution gives us some implications.
As mentioned earlier, the metric 
\begin{equation}
 ds^2 = 
e^{-2|y-y_+|/\ell}
\ga_{\m\n}dx^\m dx^\n
+ dy^2~,
\end{equation}
is a solution in the RS-II model, if $\ga_{\m\n}$ is a solution of four
dimensional vacuum Einstein equations. When we take 
the Schwarzschild metric as $\ga_{\m\n}$, 
we obtain a warped black string solution~\cite{Chamblin:1999by}. 
Although the induced metric on the brane is identical 
to the Schwarzschild black hole, this metric cannot be interpreted as 
the static final state formed after gravitational collapse 
on the brane because the solution is singular in the sense that 
the Kretschmann invariant increases indefinitely in the bulk. 
Moreover, the radius of the black string shrinks 
as we move away from the brane exponentially with 
the typical length scale $\ell$,
although the length of the black string is infinitely long. 
Therefore it suffers from 
the Gregory-Laflamme instability~\cite{Gregory:2000gf} in the region 
where the horizon radius is smaller than $\mathcal{O}(\ell)$. 

\begin{figure}[t]
\centering
\includegraphics[width=12cm, clip]{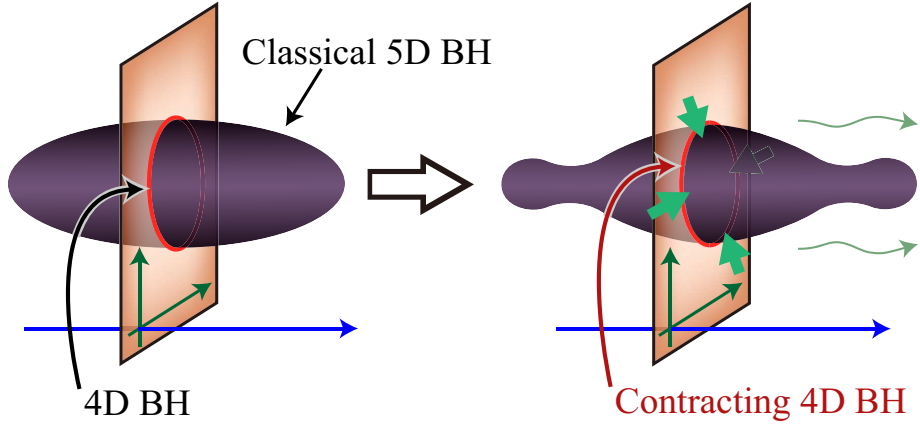}
\caption{Schematic view of the conjectured 
``classical evaporation'' process of a brane-localized black 
hole. The intersection of the five-dimensional black hole and the brane will be
 seen  by a brane observer
as a four-dimensional black hole 
with quantum correction.
Decrease in the four-dimensional black hole area due to the Hawking radiation
 would be realized in the five-dimensional perspective by the 
 deformation of the
 brane-localized black hole horizon into the bulk.}
\label{Fig:DeformingBH}
\end{figure}

From this observation on a black string, 
one may expect that the brane-localized black
hole will take a similar form to the black string chopped 
where the radius becomes $\mathcal{O}(\ell)$. 
Then, since the radius and the length of such a black object will be 
the same order at the tip, we may 
speculate that some instability similar to the Gregory-Laflamme instability 
continues to occur at the tip of this five-dimensional black hole. 
As a result of instability, 
a clump of horizon, almost pinched off from the main part 
of the horizon (see Fig.~\ref{Fig:DeformingBH}), will be formed.
Then, the clump will fall into the bulk due to the 
acceleration $a=\left(\log\sqrt{-g_{tt}}\right)_{,y}\approx 1/\ell$ 
acting universally in the bulk in the direction 
from the brane to the bulk~\cite{Gregory:2000rh}.
Such a deformation of the five-dimensional horizon in the bulk will
decrease the four-dimensional intersection between the black hole and
the brane, even though the total area of the five-dimensional black hole 
is not decreasing. 
As a result, an observer on the brane will see a four-dimensional black hole
with diminishing area.
This horizon deformation will be the five-dimensional counterpart of the 
four-dimensional black hole evaporation. This phenomenon 
is termed as classical evaporation of the brane black 
hole by Ref.~\citen{tanaka}.

Some authors consider that the black string 
is one example of a static 
quantum corrected black hole~\cite{Fitzpatrick:2006cd}.
As the induced metric on the brane is exactly identical to 
the four-dimensional Schwarzschild solution, 
there is no room for the QFT contribution to the energy-momentum tensor
in this case.
The authors of Ref.~\citen{Fitzpatrick:2006cd}
claims that this feature is due to strong coupling of the CFT, that
is, some confinement mechanism will emerge as a result of strong-coupling
effect, and hence the number of effective QFT degrees of freedom around black holes will 
be not $\mathcal{O}(N^2)$ but $\mathcal{O}(1)$. 
This reduction of degrees of freedom is assumed to be triggered by 
the curvature effects. 
Their idea is that 
the presence of small amount of curvature is sufficient 
to let most of QFT degrees of freedom inactive.  
If it is really the case, the effects of KK gravitons 
perturbatively derived on a conformally flat background, 
which includes the Friedmann universe, 
will be drastically changed once we take into account small 
gravitational wave perturbation. 
Then, many phenomena described as supporting evidence 
for the correspondence in
\S\ref{Sec:StudiesOnAdSCFTinRS} will show discontinuities 
between the cases on an exactly conformally flat background and on 
a slightly perturbed one from it, which we think quite unlikely.

There are also some other possibilities such as
\begin{itemize}
 \item the bulk/brane correspondence does not hold in the presence of black
       holes localized on the brane, and static solutions of them exist, or
 \item the bulk/brane correspondence does hold, 
but strongly coupled CFT behaves like exotic matter. 
As a result static solutions can exist.
\end{itemize}
If the former is the case, 
it would be interesting to study on the necessary condition for the
correspondence  
taking into account many examples of the correspondence we have found so far.
If the latter is the case, it will be a very interesting problem to find
the reason for absence of the
Hawking radiation. One possible mechanism to stop the radiation 
is formation of heat-insulating wall composed
of QFT. 
If such a wall really forms,
there must be some mechanism to make the thermal conductivity to be small 
since the AdS/CFT correspondence predicts very large thermal
conductivity in general~\cite{Hartnoll:b,Son:2006em}.
Another possibility is that the four-dimensional black hole on the brane is
largely deformed from an ordinary Schwarzschild black hole and the radiation is
suppressed by that. Ref.~\citen{Chamblin:2000ra}, which discusses
brane black holes charged up by contribution of the bulk Weyl tensor, may be relevant to
considerations in this line.
These issues are still open and deserve to be studied further from many aspects.

\subsubsection{Solution sequences}
\label{Sec:RS-solutionSequence}

Here, we would like to consider the phase diagram of the
five-dimensional black objects both in the ADD and RS 
models in a unified manner~\cite{Tanaka:2009zz}.
The RS-I model can be smoothly deformed from the ADD model, by 
gradually increasing bulk cosmological constant and brane tension. 
The RS-II model can be obtained by sending the negative tension brane 
to the infinity in the RS-I model. Thus,  
the phase diagram of the solution sequences in these models 
should be also smoothly connected. 
The first question to raise is how the 
phase diagram of black objects discussed in \S\ref{Sec:ADDBH}
is modified when we turn on a negative bulk cosmological
constant. A black hole floating in the bulk 
is accelerated due to the warped bulk, and
thus it cannot be static in general. 
If the black hole is sufficiently small, it can float 
near the positive tension brane
statically by balancing 
the gravitational attraction from its mirror image on the other side of the brane
and
the acceleration due to the warped bulk.
%
%
The attraction from the mirror image will not be larger than $\sim 1/R$,
where $R$ is the black hole size,
while the acceleration due to the bulk is $1/\ell$.
The former becomes too small to compete with the latter when $R\gtrsim \ell$.
Hence, we have a maximum
size for a floating black hole solution in the RS model. 
When its size is larger than that, a black hole will necessarily touch the brane
and become a brane-localized black hole.
Therefore, when the radius of the black hole is larger than
$\mathcal{O}(\ell)$, floating black holes as well as localized ones 
in Fig.~\ref{Fig:fig1_tanaka} are localized on the brane. 
This means that there are two branches of localized black holes for 
the same mass. 
\begin{wrapfigure}{l}{6.6cm}
\centering
\includegraphics[width=6.3cm,clip]{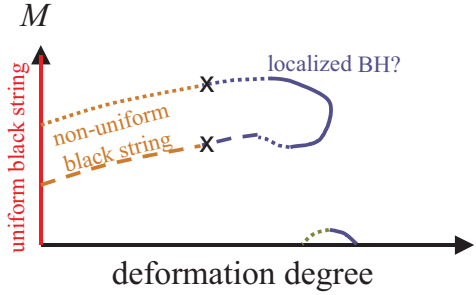}
\caption{%
Hypothesized solution sequences for the RS-I model
(This figure is taken from Ref.~\citen{Tanaka:2009zz}).
}
\label{Fig:fig2_tanaka}
\end{wrapfigure}
This situation may look a little unnatural. If we assume 
that there is no static localized black hole solution larger than 
$\mathcal{O}(\ell)$ also in the RS-I model. 
Then, to be compatible with this assumption, 
the reconnection of the sequences as shown 
in Fig.~\ref{Fig:fig2_tanaka} should take place:  
the two different sequences of a non-uniform black string will be
connected to each other forming a loop, 
while the sequence of a localized black hole is 
merged with that of a floating black hole. 

Now, the case of the RS-II model is obtained by
taking the IR brane to infinity in the RS-I model. Since the wavelength of the
lowest Gregory-Laflamme instability mode is infinitely long in this limit, the
solution sequences for the non-uniform black strings will disappear from
the solution phase diagram.
As a result, only the sequence of a brane-localized black hole 
with/without a maximum size will remain in 
Fig.~\ref{Fig:fig2_tanaka}/Fig.~\ref{Fig:fig1_tanaka}.  
Such a phase structure Fig.~\ref{Fig:fig2_tanaka} 
is consistent with the classical evaporation conjecture.

\subsubsection{Evaporation time scale and astrophysical constraints on $\ell$}
\label{Sec:astro}

Here, we study the evaporation rate of a black hole of mass $M$ due to 
the Hawking radiation through a single quantum field.
Since the Hawking radiation is approximated by black body radiation at 
the temperature $T=(8\pi G_4 M)^{-1}$, its radiation flux is given by
$F=\pi^2 T^4/60$.
Multiplying the black hole horizon area $A=4\pi (2G_4 M)^2$, 
we obtain the mass loss rate $\dot M$ as 
$\dot M = -(15360\pi G_4^2M^2)^{-1}\equiv -C/M^2$.
Solving this relation, we find that the black hole mass varies as 
$M(t)= (-3Ct)^{1/3}$. Then, the evaporation time $t_\text{evap}$ of a black hole 
with mass $M$ is estimated as 
$t_\text{evap}=M^3/{3C}= 2.10\times 
10^{67}\left({M/M_\odot}\right)^3\,\mathrm{years},$
where $M_\odot$ is the solar mass.
It implies that only the black holes lighter than $10^{-19}\Msolar\sim 
10^{11}\mathrm{kg}$ can evaporate within the Hubble time, $H^{-1}\sim 10^{10}\,\mathrm{years}$.
If we take into account the degrees of freedom of the strongly-coupled CFT, 
which is given by 
$N^2\approx \pi \ell^2 /G_4=2.36\times 10^{60}\times(\ell/14\m\mathrm{m})^2$,
the radiation power is multiplied by $15N^2\times (3/4)$ and the evaporation time 
is altered to%
\footnote{
If we omit the ``strong-coupling'' factor $3/4$
from the number of degrees of freedom, we obtain
\begin{equation}
t_\text{evap}^\text{CFT}= 
5.93 \times 10^5 \times
\left(\frac{14\,\m\mathrm{m}}{\ell}\right)^2
\left(\frac{M}{M_\odot}\right)^3\,
\mathrm{years}
= 116 \times
\left(\frac{1\,\mathrm{mm}}{\ell}\right)^2
\left(\frac{M}{M_\odot}\right)^3
\mathrm{years}~,
\nonumber
\label{lifetime2}
\end{equation}
which coincide with the values given in Ref.~\citen{Emparan:2002jp}. 
It is suggested that, however, we should include the factor $3/4$ when we 
consider a finite temperature system~\cite{Peet,Pujolas:2008rc,Murata:2008bg}.
}
\begin{equation}
t_\text{evap}^\text{CFT}
= 7.91\times 10^5\times
\left(\frac{14\,\m\mathrm{m}}{\ell}\right)^2
\left(\frac{M}{M_\odot}\right)^3
\mathrm{years}
= 155\times
\left(\frac{1\,\mathrm{mm}}{\ell}\right)^2
\left(\frac{M}{M_\odot}\right)^3
\mathrm{years}~.
\label{lifetime}
\end{equation}
We adopted $14\,\m\text{m}$ as 
the normalization of $\ell$, which is the current upper bound 
on $\ell$ obtained from table-top 
experiments~\cite{Kapner:2006si,Adelberger:2006dh,Adelberger:2009zz}.

The evaporation rate of the same order of magnitude can be explained 
also in the five-dimensional picture assuming the 
Gregory-Laflamme-type instability at the tip of the
horizon~\cite{tanaka}. 
Namely, such instability will create blobs of horizon of the size $\ell^3$
within dynamical time scale $G_4 M$, and then
the five-dimensional horizon area $\sim M^2 \ell$ will decrease as 
$\frac{d}{dt}M^2 \ell \approx \ell^3 /G_4 M$. This implies 
$\dot M\approx \ell^2/G_4 M$, which coincides with the four-dimensional estimation
of $\dot M$.

Equation~(\ref{lifetime}) relates the lifetime of a black hole with the bulk curvature
scale $\ell$. Thus, if we could give a lower bound on the lifetime of a black hole, 
we can translate it to the upper bound on $\ell$.
Since the typical black hole life time according to Eq.~(\ref{lifetime})
is very short in the cosmological sense,
we have a good chance to obtain an extremely stringent upper bound on $\ell$ 
if it
really holds.
One problem is that 
it is difficult in general
to measure the life time or its mass loss
of a black hole observationally. 


Some attempts to give bounds on $\ell$ using this conjecture
were made
by Refs.~\citen{2009ApJ...691..997J} and \citen{2009A&A...507..617J}.
They focused on X-ray binaries, and estimated evaporation rate of the black
holes from change in orbital periods of the binaries. In this way,
they obtained a bound $\ell\lesssim0.132\,\text{mm}$.
Another approach was given in Refs.~\citen{2009ApJ...705L.168G} 
and \citen{2008ApJ...683L.139Z}, in 
which the age of a black hole was estimated from observations of its host 
globular cluster. They claim that they could give a bound 
$\ell\lesssim0.003\,\text{mm}$, though there are many subtle issues in the 
observations and the logic they used to obtain this bound.
Recently, Ref.~\citen{2009arXiv0912.4744M} suggested a 
possibility that gravitational wave observations 
using LISA 
on the event rate of the extreme mass ratio inspirals
or the detection of a single galactic black hole binary 
may give an upper bound $\ell\lesssim 20\,\m\text{m}$.
Reference~\citen{Yagi:2011yu} focused on correction to the gravitational waveform
due to the black hole evaporation and usage of gravitational wave observatory
such as DECIGO~\cite{DECIGO1,DECIGO2} or BBO~\cite{BBO} which are capable of
detecting $10^5$ neutron star/black hole binaries per year.
It was shown that statistical analysis using them can give 
an upper bound $\ell\lesssim 0.33\,\m\text{m}$ after $5$ years of observation.

\subsection{Analytic approaches}
\label{Sec:RS-analytic}


We summarize analytic approaches to the black hole solutions in the RS-II model 
and the bulk/brane correspondence about it in this section.
We first mention the exact solution in the lower-dimensional model in 
\S\ref{Sec:EHM}, and then introduce perturbative approaches and 
near-horizon analysis in \S\S\ref{Sec:RS_perturbative} and \ref{Sec:RS_NH}, 
respectively. 

\subsubsection{Exact solutions in lower-dimensional model}
\label{Sec:EHM}

There is no known analytic solution of a black object in the five-dimensional
RS-II model, except for the black string~\cite{Chamblin:1999by}.
However, in the lower-dimensional RS-II model with AdS$_4$ bulk and
$(1+2)$-dimensional brane,
exact solutions of brane-localized black holes do exist~\cite{emp2}.

\begin{wrapfigure}{l}{6.6cm}
\centering%
\includegraphics[width=4.7cm,clip]{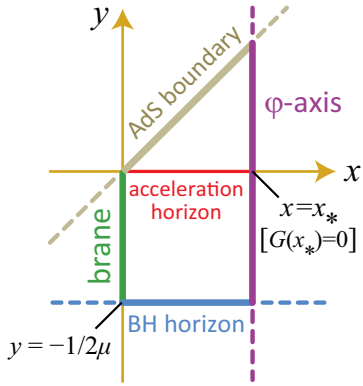}%
\caption{%
Range of coordinates of the four-dimensional 
brane-localized black hole.}%
\label{Fig:EHM}%
\end{wrapfigure}

The exact solutions are constructed from the AdS C-metric, 
whose metric is given by 
\begin{gather}
ds^2 = \frac{\ell^2}{\left(x-y\right)^2}
\left[
-Fdt^2
+ \frac{dy^2}{F}
+ \frac{dx^2}{G}
+ G d\varphi^2
\right]~,
\label{exactEHM}
\\
F = y^2+2\mu y^3,
\;\;\;
G = 1-x^2 - 2\mu x^3, 
\end{gather}
with $\mu\geq 0$.
The coordinates in this metric can be assimilated 
to ordinary polar coordinates
$(r, \theta)$ by $y\sim r^{-1}$ and $x\sim\cos\theta$.
As shown in Fig.~\ref{Fig:EHM}, the brane is 
located at $x=0$. 
An important property of this solution is that
the extrinsic curvature of the $x=0$ hypersurface 
is the induced metric multiplied by $1/\ell$. 
Therefore, cutting the solution and gluing it with its copy, we obtain a 
surface that satisfies the required junction conditions.
Thus, one can regard this plane as a $(1+2)$-dimensional brane with
constant tension.

The range of the coordinate $x$ is bounded by $x_*$, the unique 
positive real root of $G(x)=0$. The boundary at $x=x_*$ corresponds to 
the axis of $\varphi$-rotation. 
Requiring the regularity on the axis,  
the period of the angular coordinate $\varphi$ is fixed as
\begin{equation}
 \Delta \varphi= \frac{4\pi}{|G'(x_*)|}~. 
\end{equation}
Notice that $\Delta\varphi \leq 2\pi$ and the equality holds for $\mu
=0$. The black hole horizon is located at $y=-1/2\mu$. 
The AdS boundary and the acceleration horizon of the Poincare patch is given by
$x=y$ and $y=0$, respectively.
If the brane is removed, this solution describes an 
accelerating black hole solution pulled by a cosmic string in asymptotically AdS
spacetime\cite{Dias:2002mi,Krtous:2005ej}. 

This exact solution, however, is different from a brane-localized black hole in
the five-dimensional RS-II model in many regards:
\begin{itemize}
 \item 
The induced geometry on the $(1+2)$-dimensional brane, which is given by
\begin{equation}
 ds^2 =
- \left(1-\frac{2\mu \ell}{r} \right)dt^2
+ \left(1-\frac{2\mu \ell}{r} \right)dr^2
+ r^2 d\varphi^2~,
\end{equation}
does not resemble a solution in three-dimensional gravity, in which a 
massive source can create only a locally-flat conical geometry.
The appearance of a black hole solution in the present setup is 
really a four-dimensional phenomenon, and it may be
attributed to the quantum effect of CFT in the dual language.
The effective energy-momentum tensor derived from the geometry induced 
on the brane, which is given by 
\begin{equation}
 T_{\m\n}\simeq -\frac{1}{8\pi G_3}E_{\m\n}~,
\qquad
E_{\m\n}= \frac{\m \ell}{r^3}\text{diag}\left(1,1,-2\right)~,
\end{equation}
is consistent with the one generated by CFT fields with
$\mathcal{O}\left(\pi \ell^2/G_3\right)$ degrees of freedom~\cite{tanaka,emp1}
in the spacetime with a deficit angle. 
\item
The asymptotic geometry at large radii ($y\to 0$)
varies depending on the values of $\mu$. 
If we send $y$ to 0 so as not to hit the acceleration horizon, 
it should go to the point $y=0=x$. Around this point, 
the apparent $\mu$-dependent terms in the metric (\ref{exactEHM}) are all 
irrelevant. However, 
the difference of $\Delta\varphi$ from $2\pi$ persists. 
Therefore it is clear that the asymptotic structure significantly 
changes depending on $\mu$. This feature is not expected in 
the five-dimensional RS-II model. 


\item
Similarly, one can also compute 
the Kretschmann invariant of the C-metric to be given by
\begin{equation}
 R_{\m\n\ro\la}R^{\m\n\ro\la}=
\frac{24}{\ell^4}
+ \frac{48\mu^2\left(x-y\right)^2}{\ell^4}~.
\end{equation}
The first term is the contribution of the bulk cosmological constant,
and it coincides
with the Kretschmann invariant of pure AdS$_4$ spacetime.
The second term is the contribution of the black hole. The latter vanishes on the AdS
boundary $x=y$, but does not on the acceleration horizon $y=0$, having 
dependence on the azimuthal angular coordinate $x$. 
This property of remaining Kretschmann invariant on the acceleration 
horizon is not expected in the five-dimensional RS-II model.
In the five-dimensional case,
one can tell the expected form of the Kretschmann invariant 
from the Green's function given in Eq.~(\ref{GFbulk}).
It it easy to show that the Kretschmann invariant 
decays in proportion to $z^{-2}$, vanishing everywhere on
the acceleration horizon\cite{Garriga:1999yh}.
The presence of deformation beyond the acceleration horizon would mean
that the boundary condition of the problem is different for different $\mu$,
which implies that the quantum state might be also different from
the standard one in the QFT picture. Although this viewpoint has not
been widely accepted, we suspect that this will be the reason
why the Hawking radiation is absent in this spacetime 
in the dual picture, if the black hole evapolation conjecture 
is correct. 
\end{itemize}

Before moving on to the next topic, 
let us make comments on some related works. 
Ref.~\citen{Nozawa:2008wf} found linearized equations for conformal scalar,
Maxwell and massless Dirac fields on this background become separable.
Using this property, they studied quasi-normal modes for those fields,
and found that their radial part coincides with that for
four-dimensional Schwarzschild black holes while angular
part is modulated due to existence of the brane.
There were also some attempts 
to generalize the solution.
In the solution construction explained above, 
the brane tension could be detuned by changing the
parameters of the background C-metric. In this way, 
exact solutions of black holes in the KR model with asymptotically
AdS brane were constructed~\cite{Emparan:1999fd}. 
Another way of generalization is to modify 
the way of slicing the C-metric with brane. 
Such an idea was
pursued by Anber and Sorbo~\cite{Anber:2008qu}, and they constructed solutions
which can be interpreted in the four-dimensional perspective as 
a time-dependent CFT lump around a conical singularity
or two particles sustained by a string.

\subsubsection{Perturbative approach}
\label{Sec:RS_perturbative}

Another analytical approach to the brane-localized black hole is the perturbative 
approach, which applies for small mass black holes.

\begin{itemize}
 \item{\it Matched asymptotic expansion:}

First example of perturbative solutions were constructed by Karasik 
et al.~\cite{Karasik:2003tx,Karasik:2004wk}. 
When the black hole is small enough compared to the bulk curvature scale, the
spacetime near the black hole will be well approximated by a
five-dimensional Schwarzschild black hole solution. 
They perturbed this solution taking $\epsilon$, the ratio of Schwarzschild radius and 
the bulk curvature length, as a small parameter. 
The effect of brane
tension can be taken into account at the first order in $\epsilon$, and
the bulk curvature effect arises only at the second order. 
They succeeded in constructing a perturbative solution in the near-horizon region 
to the linear order in $\epsilon$, and
matched it to the perturbative solution in the asymptotic region far
from the source. 

\item{\it Accelerating black holes in AdS spacetime:}

Another perturbative approach was made by Kodama~\cite{Kodama:2008wf},
which focused on solutions describing accelerating black holes in the AdS 
spacetime. 

In this study, the gravitational field around a semi-infinite 
string gravitational source was solved perturbatively on the 
background of AdS spacetime.
As a result, 
solutions which describe black holes accelerating in asymptotically AdS
spacetime were successfully obtained.

Then, as a next step, it was assessed whether 
brane-localized black hole solutions in the RS-II model 
can be constructed
by cutting those
accelerating black hole solutions, in a similar way with the exact solution
construction in the lower-dimensional model~\cite{emp2}.
The results are:
     \begin{itemize}
      \item When the bulk dimension is five or higher, 
	    the perturbative solutions corresponding to an
            accelerating the black hole 
	    sourced by a constant tension string 
            along the symmetry axis do not admit
	    slicing with a vacuum brane.
	    
      \item 
	    When the bulk dimension is four, 
	    it is possible to slice the accelerating black hole solution
	    with a constant tension brane 
	    even when the tension of the string source is non-constant.
	    It indicates that the perturbative solutions 
            of a brane-localized black hole may have
	    degrees of freedom of a one-dimensional free function.
     \end{itemize}
If the latter is the case even in the non-linear regime,
it means that the black hole
uniqueness does not hold in the RS model, at least in the four-dimensions. 
The author gave an explanation to this breakdown of the uniqueness
from the holographic point of view:
The Einstein equations suffer from higher-derivative 
correction (see Eq.~(\ref{Z}) and below), which is the
quantum correction due to the CFT, and then the uniqueness is violated by them.


\end{itemize}

\subsubsection{Near-horizon analysis}
\label{Sec:RS_NH}

It is generally difficult to construct exact solutions of black objects.
The problem, however, can be simplified for extremal horizons,  
focusing on the near-horizon region due to symmetry enhancement. 
Such a near-horizon analysis of an extremal horizon was established by
Ref.~\citen{nearhorizon}, and the uniqueness of the near-horizon geometry
was shown. 
This analysis method was applied to the brane-localized black hole study in
Ref.~\citen{Kaus}. 

The recipient of the analysis is a brane-localized black hole which is made to
be extremal by gauge fields on the brane.
Since the surface gravity has to be constant on the event horizon to maintain 
its regularity, the event horizon in the bulk must be extremal if it is extremal
on the brane.
Focusing on this five-dimensional extremal horizon,
they successfully constructed the near-horizon geometry of an extremal
brane-localized black hole of an arbitrary size.
They confirmed that the near horizon geometry of the 
four-dimensional solution on the brane approaches the 
ordinary Reissner-Nordstrom (RN) black hole in the large black hole 
limit and evaluated explicitly the leading deviation from the RN black hole.
This analysis is interesting since, though the solution is given only in the
near-horizon region, this is the only analytic construction 
of a brane-localized black hole solution other 
than that in the lower-dimensional model~\cite{emp2,Emparan:1999fd}
we mentioned in \S\ref{Sec:EHM}.
The extension of this near-horizon analysis to the KR model case was argued by
Ref.~\citen{Suzuki}.

Note that this near-horizon analysis does not guarantee that a
static solution in the entire bulk does exist. It will be interesting to
confirm the existence of such a static solution, 
though we probably have to
resort to numerical technique similar 
to Refs.~\citen{Wiseman:2001xt,Kudoh:2003xz,Kudoh:2004kf,Yoshino:2008rx}
to construct it.

\subsection{Numerical approaches}
\label{Sec:RS-numerical}

In this section, we summarize numerical approaches to the issues of 
brane-localized black holes in the RS-II model. 
After mentioning some approximative approaches 
in \S\ref{Sec:RS_numerical_approx},
we explain the current status of numerical solution construction 
in \S\ref{Sec:RS_numerical_direct}.

\subsubsection{Approximative approach}
\label{Sec:RS_numerical_approx}

Since direct approaches toward numerical solution
construction is rather difficult, 
some approximative approaches for solution
construction have been developed. 
One of such approximative approach was discussed in Ref.~\citen{Creek:2006je}.
Their trick is to use a known exact black hole solution for the bulk spacetime
metric, and consider the brane trajectory on such a background.
One of the background bulk spacetime they used is the five-dimensional AdS
Schwarzschild spacetime. Since this bulk does not admit vacuum brane
slicing, they put some matter on the brane to force the brane to be static. In this
way, they constructed a static solution of a brane-localized black hole in the
RS-II model with some artificial matter on the brane. 

Another approach is to relax the static condition.
By requiring the system to be only momentarily static, instead of being
completely static, we can find a vacuum brane slicing on the AdS-Schwarzschild
bulk only by solving a set of ordinary differential equations. 
Such a solution might inherit some properties of static solutions. 
Static
solutions should be contained in the whole set of momentarily-static
solutions, and
we may expect that a solution with maximum entropy in a whole set of 
momentarily-static solutions to is a static solution.
Based on this idea, Ref.~\citen{Tanahashi} studied the properties of
momentarily-static solutions of brane-localized black holes.
Although the investigated momentarily-static solutions 
were very limited, 
the results were consistent with the classical black hole conjecture of
the brane-localized black holes.
An extension of this approach 
to the KR model was discussed in Ref.~\citen{fBH}.

\subsubsection{Direct approach to the solution construction}
\label{Sec:RS_numerical_direct}

First attempt of numerical solution construction was made 
by Kudoh et al.~\cite{Kudoh:2003xz,Kudoh:2004kf}
based on the numerical technique explained in \S\ref{Sec:ADDBH}. 
Solving a set of three elliptic equations numerically by the relaxation method,
solutions for small static brane-localized black holes were
successfully constructed. However, their numerical code failed 
to find solutions with the horizon size larger than $\sim 0.2 \ell$.
Kudoh~\cite{Kudoh:2004kf} applied this numerical technique to
the brane-localized black holes in six-dimensional RS-II model, and constructed
solutions whose size is as large as $\sim 2 \ell$.

This numerical analysis was re-analyzed by Yoshino~\cite{Yoshino:2008rx}.
The improvements are in the choice of coordinates and variables as well 
as in the accuracy of numerical differentiation.
In Ref.~\citen{Kudoh:2003xz}, they used 
the coordinates $(\rho,\xi)$ defined from Eq.~(\ref{conformalAnsatz}) by 
\begin{equation}
 r= \rho\sin\chi~,
\qquad 
z= \ell + \rho\cos\chi~,
\qquad
\xi = \chi^2~.
\end{equation}
The angular coordinate $\xi$ was employed to eliminate terms proportional to
$1/\sin^2\chi$, which becomes singular on the axis of rotational symmetry, 
from the equations. 
Reference~\citen{Yoshino:2008rx} used coordinates $(x, \chi)$ defined by 
$x\equiv\log(\rho/\rho_\text{horizon})$ instead.
$\chi$ was chosen to increase the number of data points near
the axis,
and $x$ was introduced to let the asymptotic boundary, at which the
metric is set to coincide with the unperturbed RS model, 
much further compared to Ref.~\citen{Kudoh:2003xz}.
One drawback of this choice is the re-appearance of singular term 
$\propto 1/\sin^2\chi$. 
In Ref.~\citen{Yoshino:2008rx} these singular terms were 
eliminated as far as possible by choosing appropriate 
linear combinations of the free functions $R$ and $C$ 
as variables to solve. 
About the accuracy of differentiation, 
Ref.~\citen{Kudoh:2003xz} used a scheme
with second-order accuracy while Ref.~\citen{Yoshino:2008rx} used one 
with fourth-order accuracy in the bulk and third-order on the boundaries.

As a result of this numerical analysis, in Ref.~\citen{Yoshino:2008rx}
Yoshino claimed that he observed 
``nonsystematic error'', which is the error distinct from systematic numerical errors
due to, e.g., finite grid size or finite distance between the asymptotic boundary and the
black hole, and it dominates when the asymptotic boundary is moved toward infinity.
Based on this observation, 
it was 
guessed that even small static brane-localized black holes may not exist.
Ref.~\citen{Kleihaus:2011yq} studied brane-localized black holes charged with
respect to a Maxwell field on the brane, and achieved similar results
for non-extremal black holes.
However, we think that 
the results in Ref.~\citen{Yoshino:2008rx} 
are not inconsistent with the
existence of small static black holes. When the grid size is 
reduced, the error reported in Ref.~\citen{Yoshino:2008rx} 
looks always diminishing for any fixed location of the asymptotic boundary. 

{\it Note added}: 
After finishing this review, a paper by 
Figueras, Lucietti and Wiseman appeared~\cite{Figueras:2011va}.
The paper uses numerical technique based on a version of 
Ricci-flow equation to solve the static Einstein
equations~\cite{Headrick:2009pv}, 
and reports that there is a static black hole solution attached to the 
AdS boundary with the induced metric is conformal to four-dimensional
Schwarzschild black hole. 
This means that CFT coupled with
four-dimensional gravity does not radiate even if there is a black hole
at the leading order of $1/N$ expansion.
Numerical results of their subsequent paper
implies that this solution is smoothly related to 
brane-localized black hole solutions in the RS-II model~\cite{Figueras2}.

Here we explain our tentative understanding of this result. 
The result shows that the
energy density of CFT increases near the horizon to the level 
comparable to the radiation fluid composed of $O(N^2)$ 
degrees of freedom at the Hawking temperature. 
In this sense one may say that the pressure of CFT cloud is 
in balance with the pressure of the Hawking radiation from 
the black hole horizon. 
This increase of energy density near the horizon could 
be expected to some extent 
by extrapolating the energy density around a point particle, 
which behaves like $\propto N^2 r_g/r^5$, 
where $r_g$ is the gravitational 
radius of the point particle. Near the horizon, it blows up 
to $N^2/r_g^4\approx N^2 T_{BH}^4$.  
The behavior of the CFT cloud around a point particle,
which is similar to that for their black hole solution in asymptotic region,
however,
is far different from radiation fluid. 
The pressure is comparable to the energy density, 
decaying proportional to $1/r^5$, but 
it is not isotropic at all. Tangential pressure is negative and 
$3/2$ times bigger in magnitude than the radial one. 
The force balance in the radial direction is established 
essentially between the radial pressure gradient and 
the tension in the tangential directions. 
We should note that
all of these features can be understood also in terms 
of the language of weakly coupled CFT. 
Such CFT contribution 
to be attributed to the vacuum polarization caused by spacetime 
curvature would be
distinguished from thermal component as in the case of dark 
radiation. In the latter case one can freely change the amplitude 
of CFT energy density, and it seems to behave like radiation fluid.  

Many new questions arise to obtain a consistent picture to 
understand the behavior of strongly coupled CFT. 
The evidences previously 
supporting the black hole evaporation conjecture as mentioned 
in this section might be now in conflict with the
existence of static black hole solutions. 
The arguments were mostly based on the analogy of the weakly coupled 
CFT or radiation fluid. We need to clarify which part of the 
analogy breaks down. 

\section{Black holes in other braneworld models}
\label{Sec:other}
\subsection{Black holes in KR model}
\label{Sec:KR}

In the KR model, the geometry on the brane is AdS$_4$. 
Let us firstly remind basic 
properties of black holes in equilibrium with 
radiation in asymptotically AdS spacetime.
The negative cosmological constant of the AdS spacetime exerts contracting
force to fields in it. In other words, 
the effective gravitational 
potential rises up to infinity at the AdS boundary, 
and it effectively plays as a bounding box. 
Then, a quantum black hole in the asymptotic AdS spacetime can 
be in thermal equilibrium with its Hawking radiation since 
the emitted radiation is trapped by this gravitational potential well.
Furthermore, it is known that 
black holes larger than the curvature scale of AdS spacetime 
have positive specific heat and thus they are thermodynamically stable.
In contrast, black holes smaller than AdS curvature scale 
evaporate completely without reaching thermal 
equilibrium. 
This implies that there is a phase transition between thermal AdS phase at
low temperature and the AdS black hole phase at high temperature. 
Such phase transition in the canonical or micro-canonical ensemble 
is known as the Hawking-Page transition~\cite{HP}.

Keeping in mind the above-mentioned 
properties of black holes in asymptotically AdS spacetime, 
let us consider the implication of the bulk/brane correspondence 
on brane-localized black holes in the KR model.
(What we consider here seems to be different blanch of solutions 
that can exist also in RS-II setup. In that case the energy density of
CFT vanishes at infinity and is far from the thermal one.)
First of all, the bulk/brane correspondence implies that there should be 
a large static brane-localized black hole as its five-dimensional 
counterpart.
It also predicts that there should be
some transition in the bulk spacetime which is dual to the Hawking-Page 
transition. 
Such a possibility was firstly considered in Ref.~\citen{CK}, which claimed that the 
bulk counterpart of the Hawking-Page transition is given by
the transition between thermal AdS$_5$ phase and the 
five-dimensional AdS black string phase.
Since the projection of the AdS black string solution onto the brane is 
four-dimensional AdS Schwarzschild spacetime, this five-dimensional 
phase transition is quite in parallel to that in four dimensions.

A five-dimensional black string in AdS spacetime, however, requires infinite 
energy to be produced from the thermal bath since 
the black string extends to the AdS boundary and the horizon 
area and the mass diverge there.\footnote{
Gregory et al~\cite{Gregory:2008br}.~focused 
on the fact that a black string has a finite minimum radius
at the ``throat'' of the bulk. 
If the horizon radius on the brane is 
larger than the four-dimensional AdS curvature length, 
the minimum radius in the bulk is larger than 
the five-dimensional curvature length and then the
black string does not suffer from Gregory-Laflamme instability. 
Therefore they considered that the black string is the 
stable final state of gravitational collapse on the brane  
in this case. 
The effective energy-momentum tensor of the QFT derived 
from the induced metric is obviously proportional to the
induced metric. Namely, the QFT
contributes only to the renormalization of the 
four-dimensional cosmological constant. 
They also computed the energy-momentum tensor due to QFT in the weak coupling 
approximation, and confirmed that it is not proportional to the metric
at all. 
They interpreted this discrepancy to be the effect of strong coupling.
However, we think that 
black string cannot be the final state after gravitational collapse on 
the brane. 
}
Such an infinite energy input is possible in a canonical ensemble with a fixed 
temperature, but impossible in a micro-canonical ensemble with 
a fixed total energy.
In the actual dynamical gravitational collapse, the amount of 
energy that can be used is limited. Hence, formation of infinitely long 
black string is prohibited. 

Another scenario of the transition in five dimensions was proposed 
in Ref.~\citen{Tanaka:2009zz}
taking the following points into account.  
\begin{itemize}
\item 
The holography implies the
existence of a static brane-localized black hole larger than the
four-dimensional bulk curvature scale on the brane.

\item
Since the KR model can be obtained by smooth deformation of the RS
 model, i.e., by
decreasing the brane tension, the phase diagram of black objects 
in the KR model must be smoothly connected to that in the RS model. 

\item
Since black string solutions cannot be obtained with a finite 
amount of energy, they are not included in the phase diagram 
of solutions as long as 
equilibrium states in the micro-canonical ensemble 
are concerned. 
\end{itemize}

\begin{wrapfigure}{r}{6.5cm}
\centering
\includegraphics[width=6.5cm,clip]{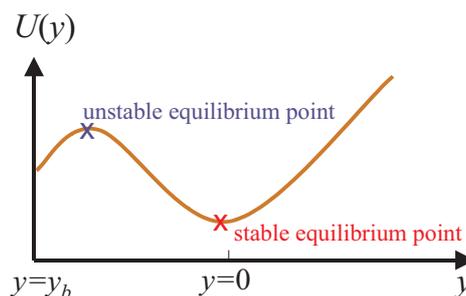}
\caption{%
Effective potential for a small particle in the KR model bulk.
(This figure is taken from Ref.~\citen{Tanaka:2009zz}.)
}
\label{Fig:fig3_tanaka}
\end{wrapfigure}

To consider the bulk black holes in the KR model,
let us first examine properties of a particle placed in the bulk of this model.
It is convenient to introduce the coordinates given by
\begin{equation}
ds^2 = dy^2 + \ell^2 \cosh^2\left(y/\ell\right)ds^2_{\text{AdS}_4}~,
\end{equation}
where $ds^2_{\text{AdS}_4}$ is the metric of AdS$_4$ with unit curvature.
The particle placed in the bulk will feel acceleration due to the bulk 
cosmological constant, 
and its value will be given by
$a=\left(\log\sqrt{-g_{tt}}\right)_{,y}$. 
Then, the effective gravitational potential due to the warped bulk 
geometry will be given by 
\begin{equation}
 U_\text{eff} = \log\left(\sqrt{-g_{tt}}\right) 
= \log\left\{ \ell \cosh\left(\frac{y}{\ell}\right)\right\}~.
\end{equation}
In addition to that, the gravitational attraction between the particle and its
mirror image will also contribute to the effective potential.
The total effective potential for a small particle in the bulk 
will be schematically as shown in Fig.~\ref{Fig:fig3_tanaka}.
This effective potential will have two extrema
at which the small particle can stay at rest: a stable minimum at the
``throat'' of the bulk and an unstable maximum near the brane.

Now we will consider black hole solutions in the bulk. When the black hole is
sufficiently small, it will be approximated simply by a point particle. 
Then, the effective gravitational potential for this small black hole will be
similar to that for a point particle shown in Fig.~\ref{Fig:fig3_tanaka}.
Corresponding to the two extrema in the effective potential,
there will be two solutions for such a small black hole for a fixed mass,
and thus there will be two sequences of solutions of a floating small black
hole.
One sequence represents a black hole staying at the unstable maximum near the
brane, and the other represents the one staying at the stable minimum at the
``throat'' of the bulk.

\begin{wrapfigure}{r}{6.6cm}
\centering
\includegraphics[width=6.5cm,clip]{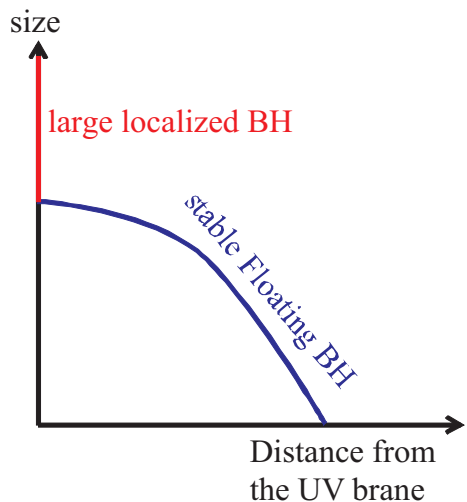}
\caption{%
Proposed phase diagram of the black hole solution sequences in the KR model.
}
\label{Fig:fig4_tanaka}
\end{wrapfigure}
To get a perspective on the entire solution sequences, let us consider 
what will happen when we increase the black hole mass.
First, we study a black hole at the unstable maximum. Since that black hole
is sustained by gravitational attraction from its mirror image
behind the brane, its properties will be similar to those of a floating black hole 
in the RS-II model (see \S\ref{Sec:RS-solutionSequence}). 
Next, we study a black hole at the stable minimum, which is located at the 
``throat'' of the bulk.
Such a black hole will sit there until it becomes sufficiently large to fill up
the region between the ``throat'' and the brane.
Increasing further the black hole mass, 
the black hole will touch the brane and then it will become a brane-localized 
black hole.
We summarize the latter solution sequence 
deduced from the above 
consideration in Fig.~\ref{Fig:fig4_tanaka}.

Let us consider further 
how the solutions in the latter sequence look like on the brane and 
the stability of those solutions in the five-dimensional point of view.
When the black hole larger than the bulk curvature scale
is floating near the ``throat'' in the bulk, 
the temperature of the black hole increases as we increase the 
mass of the black hole. (Notice that the surface gravity of 
an AdS black hole increases as its mass becomes bigger when 
the size is bigger than the AdS curvature length.)
In the four-dimensional dual picture, 
the energy-momentum tensor due to QFT will be approximated by 
that of 
radiation fluid with a large number of degrees of freedom in 
thermal equilibrium at the black hole temperature, 
and hence the energy density will 
increase as the black hole becomes bigger. 
At some critical size, the bulk black hole will touch the brane. 
The intersection of the five-dimensional black hole 
with the brane will be observed 
from a brane observer as a four-dimensional black hole 
associated with surrounding radiation.
Such a configuration with a large five-dimensional black hole 
and a small four-dimensional intersection looks unstable.
By increasing the five-dimensional black hole size further, the size 
of the four-dimensional black hole will become larger than the four-dimensional 
AdS curvature scale. In this regime, the solution will look like the 
black string solution truncated at around the ``throat''. 
Such a five-dimensional black hole with a large four-dimensional
intersection will be stable. 

Assuming the bulk/boundary correspondence, we can estimate the physical 
parameters at the critical points.
Firstly, the size of the five-dimensional black hole when it touches 
the brane is estimated as follows.
When the five-dimensional black hole touches the brane, the temperature at the 
touching point will diverge since local temperature is proportional to the 
red-shift factor $1/\sqrt{-g_{tt}}$. 
In the four-dimensional dual picture, 
this configuration would be described by 
self-gravitating radiation fluid whose density diverges at the center. 
Properties of such radiation fluid star in asymptotically AdS spacetime
was studied by Ref.~\citen{PP}, and it was shown that such a star 
with singular central density has entropy of 
$\mathcal{O}\left(G_4^{-1}\ell^{1/2}L^{3/2}\right)$.
This entropy should coincide with the five-dimensional entropy 
$S\sim A_5 / G_5\sim A_5 / \left(\ell G_4\right)$ 
if the bulk/brane correspondence holds. Assuming so, we find that the 
five-dimensional black hole area and the five-dimensional black hole 
radius at this critical point to be 
$A_5\sim \left(\ell L\right)^{3/2}$ and $r_\text{5DBH}\sim (\ell L)^{1/2}$,
respectively.

Next, we estimate the size of a large brane-localized black hole at 
the transition of stability. 
Again, in the dual four-dimensional picture, 
a black hole is stable when its mass 
becomes comparable to the total radiation energy around 
the black hole, $N^2T^4L^3\sim G_4^{-1}\ell^2L^3T^4$.
Then, the critical temperature will be given by 
$T\sim (\ell^2L^3)^{-1/5}$, where we used $M\sim (G_4 T)^{-1}$ 
assuming $T\ll L^{-1}$. 
Then, the critical mass and horizon radius will be given by 
$M\sim G_4^{-1}(\ell^2L^3)^{1/5}$ and $r_\text{4DBH}\sim(\ell^2L^3)^{1/5}$,
respectively. 
This $r_\text{4DBH}$ will be the critical radius of 
the intersection of the brane and the brane-localized black hole.
In Ref.~\citen{Kashiyama}, the above estimates 
were conducted more explicitly 
in the four-dimensional point of view
using radiation fluid approximation, 
taking into account the back reaction to the geometry.  

These expectations about the solution sequences was studied in 
Ref.~\citen{fBH} from the five-dimensional point of view.
In this work, time-symmetric initial data of five-dimensional black hole floating in the
bulk of the KR model was constructed, and its properties, such as
thermodynamic stability or effective energy density profile induced on the
brane, were studied. The results basically supported
the arguments presented above based on the bulk/brane correspondence. 

{\it Note Added}:
Here we assumed that the black holes are in thermal equilibrium 
with the CFT cloud. This expectation might be wrong from the fact 
that there are brane-localized black hole solutions in RS-II model, 
which are not in equilibrium in the sense that the asymptotic energy 
density of CFT vanishes. 
However, even in that case one can argue that 
local pressure balance between the 
black hole radiation and the surrounding cloud is satisfied. 
Furthermore, the existence of the sequence of solutions discussed 
in this subsection is 
naturally expected from the consideration in five-dimensional gravity 
side without relying on the bulk/brane correspondence.  

\subsection{Black holes in DGP model}
\label{Sec:DGPBH}

The perturbative analysis of gravity in the DGP model 
requires a special care since a naive perturbative
expansion around a pure four-dimensional solution breaks down in the short scale
$r\ll r_*$. Thanks to some studies, it was noticed that it is the brane bending which
becomes non-linear in this range, and that the perturbative analysis can be
accomplished only by taking the second order perturbation with respect 
to the brane bending at least when matter source is sufficiently smoothed out.
This analysis indicates that the gravitational field around a source of
four-dimensional mass $M$ is roughly given 
as summarized in
Table~\ref{Tab:perturbativeDGP}.
\begin{table}[t]
\caption{Gravitational field of mass $M$
in the DGP model clarified by a perturbative
 analysis.}
\centering
\begin{tabular}{|l|l|}
 \hline
 $r_g \ll r \ll r_*$ &  
 four-dimensional Schwarzschild solution for mass $M$
 \\ \hline
 $r_* \ll r \ll r_c$ &  
     solution interpolating four and five-dimensional solutions
     \\ \hline
 $r_c \ll r$ & 
     five-dimensional Schwarzschild solution for mass $M$\\
 \hline
\end{tabular}\vspace{-6.5mm}
\label{Tab:perturbativeDGP}
\end{table}

A perturbative analysis, however, does not apply to general matter source
and breaks down especially in region $r\sim r_g$ where the solution 
will become non-linear.
To investigate the gravity in this region, we have to solve the
Einstein equations non-perturbatively to construct a black hole solution.
This direct approach is not accomplished so far. 

An approximative 
method that simplifies the problem is to introduce an {\it ad-hoc} metric ansatz. 
Such an attempt was done in 
Ref.~\citen{Middleton:2003yr} and approximate solutions including 
the region $r\sim r_g$ were obtained. 
Using another ansatz $-g_{tt}=1/g_{rr}$ on the brane,
the authors of Ref.~\citen{G-Sch}
extracted a closed equation of the lapse function on the brane from the full
Einstein equations.
By solving this equation, they fixed
the induced metric and a part of extrinsic curvature on the brane.
The difference of their result
from the perturbative result summarized 
in Table~\ref{Tab:perturbativeDGP}
is in the long range region $r\gg r_*$. Their solution converges into a
five-dimensional Schwarzschild spacetime not with $M$ but 
with a screened mass $M\left(r_g/r_c\right)^{1/3}$.  

An alternative approach to study the gravitational field in the DGP model is to
use the shockwave solutions, 
which describes a gravitational field induced by a
particle on the brane moving in a relativistic speed.
One advantage of this approach is that exact solutions in the entire bulk 
can be obtained by rather simple analysis.
This analysis showed that the interpolation between four-dimensional behavior in
short scale and five-dimensional one in large scale does persist even in the
presence of dynamical gravity generated by
Planckian scattering of particles~\cite{DGPShock1,DGPShock2}.
This approach may provide a new window to non-linear gravitation in the DGP model.
\vspace{-1.5mm}
\section{Summary and outlook}
\label{Sec:summary}
\vspace{-1.5mm}
In this review,
we introduced basics about representative 
braneworld models and their gravitational properties,
and summarized current understandings about black hole solutions in those models.
Perturbative gravity is well understood in these braneworld models, 
but black hole solutions are less known yet. 
In particular, explicit numerical construction of solutions in the
models with a warped extra-dimension such as the RS model and the KR model 
is still very limited. 
We therefore invoked arguments based on the bulk/brane correspondence in braneworld
models, to predict the features of black hole solutions in such models.

The primary interest in this context would be 
to prove the bulk/brane correspondence in braneworld
models, which seems difficult to accomplish in a formal way, though. 
The bulk/brane correspondence in braneworld 
naively predicts classical evaporation of 
a brane-localized black hole in the
RS-II braneworld model. This prediction, however,   
{\it very recently} seems to be denied by a 
beautiful numerical study by 
P.~Figueras, J.~Lucietti and T.~Wiseman, although further 
confirmation by independent calculation is wanted. 
The solution they have provided indicates 
that CFT can be in equilibrium with 
a black hole without emitting Hawking radiation 
at the leading order of the so-called large $N$ expansion. 
This proves that this field of research is still very active, 
and that this review is really written 
in the middle stage of rapid growth. We hope that this review 
provides a guide for further development. 

Studies of black holes in the modified setup, such as braneworld models 
with bulk Gauss-Bonnet term or the DGP braneworld model, would be also 
very fruitful, though such studies 
are rather limited at this moment compared to those for the RS model. 
It would be interesting to study the properties of black hole solutions 
or the holography in these modified settings, too.

\section*{Acknowledgements}
We would like to thank Nemanja Kaloper for useful comments on the manuscript.
The Japanese Society for Promotion of Science Grants N.~21244033, 
the Global COE Program ``The Next Generation of Physics, Spun from Universality
and Emergence'', and the Grant-in-Aid for Scientific Research on Innovative
Areas (N.~21111006) from the MEXT are gratefully acknowledged for their support.
NT is supported by the DOE Grant DE-FG03-91ER40674.

\end{document}